\documentclass{ws-procs975x65}            
\usepackage{color}
\usepackage[colorlinks,citecolor=blue,urlcolor=blue]{hyperref}
\renewcommand{\vec}[1]{\boldsymbol{#1}}
\def \D{\Delta}

\def \k{{\vec{k}}}

\def \r{{\vec{r}}}

\def \K{{\mathbf{K}}}
\def \p{{\vec{p}}}
\def \Q{{\vec{Q}}}
\def \tn{\textnormal}

\def \e{\varepsilon}

\def \beq{\begin{eqnarray}}
\def \eeq{\end{eqnarray}}
\def \vp {\varphi}
\def \ve {\varepsilon}
\def \v {\vec{v}}
\def \L {{\cal{L}}}
\def \n {{\bf n}}

\begin{document}

\title{The enigma of the pseudogap phase of the cuprate superconductors}

\author{Debanjan Chowdhury}

\address{Department of Physics, Harvard University, \\
Cambridge Massachusetts-02138, U.S.A. \\
 debanjanchowdhury@gmail.com}

\author{Subir Sachdev}

\address{Department of Physics, Harvard University, \\
Cambridge Massachusetts-02138, U.S.A, and,\\
Perimeter Institute of Theoretical Physics, \\
Waterloo Ontario-N2L 2Y5, Canada. \\
sachdev@g.harvard.edu
}

\begin{abstract}
The last few years have seen significant experimental progress in 
characterizing the copper-based hole-doped high temperature superconductors
in the regime of low hole density, $p$. Quantum oscillations, NMR, X-ray, and STM
experiments have shed much light on the nature of the ordering at low temperatures.
We review evidence that the order parameter in the non-Lanthanum-based cuprates is a $d$-form factor density-wave.
This novel order acts as an unexpected window into the electronic structure 
of the pseudogap phase at higher temperatures in zero field: we argue in favor of a `fractionalized 
Fermi liquid' (FL*) with 4 pockets of spin $S=1/2$, charge $+e$ fermions enclosing an area 
specified by $p$.
\\~\\
{\tt Proceedings of the 50th Karpacz Winter School of Theoretical Physics,\\ 2-9 March 2014, Karpacz, Poland}
\end{abstract}

\keywords{Hole-doped cuprates, Pseudogap, Bond-density wave, Fractionalized Fermi-liquid.}

\bodymatter

\section{Introduction}
\label{intro}
\begin{figure}
\begin{center}
\includegraphics[width=4.5in]{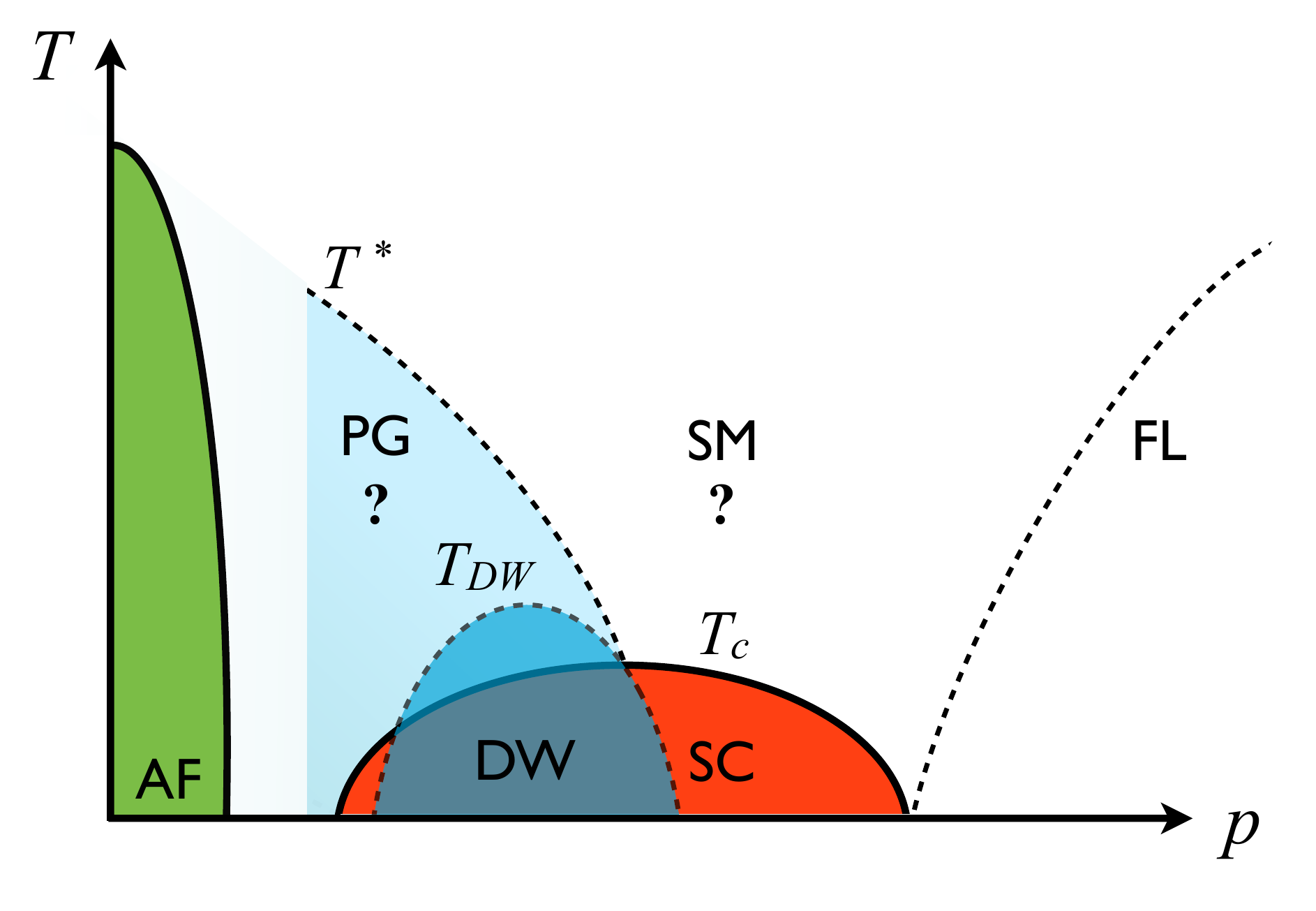}
\end{center}
\caption{A temperature ($T$) vs. hole-doping ($p$) phase diagram for the non-La-based cuprates. The parent state is an antiferromagnetic (AF) Mott insulator which upon being doped leads to a $d$-wave superconductor (SC) with transition temperature, $T_c$. The extremely overdoped phase is a familiar Fermi-liquid (FL). The two mysterious phases (denoted `?')  are the pseudogap (PG), which onsets at a high temperature ($T^*$), and the strange-metal (SM), occupying a wide fan-shaped region. Multiple recent experiments have investigated the nature of the lower ($T_{DW}<T^*$) temperature density-wave (DW) phase in the underdoped cuprates, which will be a central topic of this review. }
\label{pd}
\end{figure}

 Ever since their discovery \cite{BM86} in 1986, the cuprate high-temperature superconductors (SC) have been among the most 
 well-studied correlated electron materials, both theoretically and experimentally. These materials show a whole zoo of different phases
 as a function of hole doping. The parent antiferromagnetic Mott insulator, the $d$-wave superconductor and the Fermi liquid state at large doping are reasonably well understood at this point. On the other hand, the phases that continue to elude an explanation, and occupy a large portion of the phase diagram (Fig.~\ref{pd}), are the `pseudogap' and the `strange-metal'. It is essential to develop a microscopic understanding of these phases
in order to resolve the mystery of cuprate superconductivity, as they are the `normal' states out of which superconductivity (and various other ordering tendencies) arise at low temperatures. 

Although we don't have a complete understanding of these phases, a lot is now known phenomenologically about the pseudogap regime of the underdoped cuprates. The first signatures of the `high'-temperature pseudogap ($T<T^*$) came from measurements of the Knight-shift \cite{HA88}, followed by other spectroscopic measurements \cite{ZXSRMP03}, which showed evidence for a suppression in the density of states at the Fermi-level along specific regions of the Brillouin zone. However, there is no clear evidence for any broken symmetry in this regime. On the other hand, in recent years many remarkable experiments have revealed the nature of the `low'-temperature pseudogap ($T<T^{**}<T^*$), where a number of broken symmetries have, in fact, been observed. A major breakthrough was the initial observation of quantum oscillations at high magnetic-fields and very low temperatures, followed by the subsequent discovery of short-ranged charge-order even at zero field in the underdoped regime. This was marked by initial optimism that, perhaps, the ``hidden" broken-symmetry phase responsible for giving rise to the pseudogap has finally been identified. However, a more systematic study has now made it quite clear that the somewhat fragile charge density wave itself can't be responsible for causing the pseudogap; but it possibly arises at low temperatures out of the {\it normal} pseudogap phase with no broken symmetry (there may be broken discrete symmetries, 
such as time-reversal or nematicity \cite{FKT14}, at higher temperatures, but these have little direct effect on the electronic structure or spin excitations). 
One might then ask, what have we really learnt post the discovery of charge order? Could it perhaps serve as a ``window" into the nature of the pseudogap state itself?

In this review, we shall take the point of view that, indeed, the {\it micro-structure} of the charge-order, namely its `form-factor' and (doping-dependent) wavevector, shall provide us important clues about the nature of the higher temperature pseudogap phase. We will investigate low-energy particle-hole instabilities of possible normal states, and this will help us identify the necessary features required to give rise to a state that is qualitatively similar to the one observed in the experiments. This will also aid us in distinguishing between some of the different approaches that have been used to describe the pseudogap phase.
 
Broadly speaking, theoretical approaches for characterizing the pseudogap can be classified into two major categories. In the first category, the build up of antiferromagnetic correlations and the opening of a spin-gap \cite{HA88} at $T^*$ signal the onset of a quantum spin liquid \cite{LNW06,YRZ06, LVVFL12,MPSS12,MV12}. Moreover, as a function of decreasing temperature, there could be multiple crossovers within the metallic spin-liquid and there could also be instabilities to other symmetry-broken phases{\footnote{Though clearly important for describing some of the phenomenology, we shall ignore the effects arising due to the presence of {\it quenched disorder}, which in some cases forbids a true symmetry breaking in two spatial dimensions.}} at lower temperatures. However, in order for this to be a useful and precise characterization, there should be remnants of the `topological order' of the spin liquid at high 
temperatures{\footnote{By `topological order', we mean states with emergent gauge excitations; for states with an energy gap, there
are non-trivial ground state degeneracies on a torus.}}. One possibility is the presence of closed Fermi pockets which violate the Luttinger theorem constraining the total area enclosed by the Fermi surface \cite{PJ11}, and this may be related to photoemission spectra which have intensity only 
on open arcs in the Brillouin zone. 
In this approach, the pseudogap metal found at high temperatures is a novel quantum state which, with moderate changes, could be stable at low 
temperatures for suitable model Hamiltonians.

In the second category \cite{SKRMP03,SSRMP,JSDP98,MV09}, the antiferromagnetic correlations are precursors to the appearance of antiferromagnetism, superconductivity, charge density-wave, and possibly other conventional orders at low temperatures. In the pseudogap regime, we then have primarily thermal and classical, rather than quantum, fluctuations of these orders. This raises an immediate question: why don't other unconventional superconductors, many of which have a robust antiferromagnetic phase \cite{DSRMP12}, also display pseudogap behavior due to the precursor thermal fluctuations? What is so different about the hole-doped cuprates? 
 
 In this review, we shall delve into discussing some of the merits of both approaches. However, hopefully by the end of this article, it will become clear to the reader that we believe that the spin liquid approach 
 provides a more coherent picture of the hole-doped side of the phase diagram. 
 
Let us now outline the scope of this review. There are already a number of excellent reviews that discuss much of the previously known phenomenology of the pseudogap phase \cite{MRCK05,SKRMP03,LNW06}. Our emphasis here will be on the more recent experiments, particularly with regard to the discovery of charge order in the non-Lanthanum-based cuprates, and the implications of these experiments on the nature of the high temperature pseudogap phase. We shall not explicitly consider experimental reports of nematic order \cite{FKT14}: this order can be an ancillary consequence
of our low and high temperature models of the pseudogap, but we do not believe it is crucial to the basic phenomenology.
Also, we shall not discuss any of the signatures of time-reversal symmetry-breaking that have possibly been seen via polarized neutron scattering \cite{PB08} and Kerr-rotation \cite{AK08} experiments, but surprisingly not in $\mu$SR or NMR experiments. While these are clearly interesting experiments that need a theoretical explanation in the future, we shall choose not to discuss them any further in this review.

With the above goals in mind, let us raise a few sharp questions with regard to the pseudogap phase, and the charge density wave in particular, that we hope to address in the rest of this review: 
\begin{itemize}
\item What is the normal state out of which the charge density wave emerges? In other words, can the charge density wave help us identify the nature of the parent pseudogap state?
\item What is the modulation wavevector, and, the internal symmetry (`form-factor') of the charge density wave?
\item What is the nature of the onset, and in particular, the temperature dependence of the charge density wave correlations? 
\item Does the charge density wave lead to Fermi surface reconstruction that is consistent with quantum oscillations at high fields, and if so, which underlying Fermi surface does it reconstruct?
\item Is there an underlying quantum-critical point (QCP) that controls the physics of the strange-metal region, and if so, what are the two phases on either side of this QCP?
\end{itemize}

The reader will probably agree that most of these questions, while easy to state, might not have a complete answer at present. However, we shall argue that based on a body of work done over the past few years, it is at least partially possible to answer some of them. Some of the above questions will necessary require us to speculate, with the hope that future work will help us address these issues further. 

The rest of this review is organized as follows: In Section~\ref{expts}, we summarize some of the recent path-breaking experiments that have helped elucidate the nature of the charge-order in the underdoped cuprates. We also define a convenient way of expressing the modulation of the charge density in terms of a bond-observable. In 
Section~\ref{pheno}, we shall describe the phenomenology associated with the low temperature pseudogap in terms of fluctuating orders at zero field and long-range charge-order at high-field, without reference to any particular microscopic theory. In Section~\ref{flstar}, we shall depart from this `classical' perspective and consider a theory for a `fractionalized' Fermi-liquid (FL*) in the high temperature pseudogap. This will involve two alternative routes towards arriving at a description of doped carriers moving in a background spin-liquid. In 
Section~\ref{largeFS}, we shall attempt to connect the low and high temperature descriptions by first reviewing the instabilities of a Fermi-liquid interacting via antiferromagnetic (AFM) interactions. This approach will lead to results that are only partly consistent with the experiments, 
but will give crucial insight into the structure of the problem. However, it will serve as a nice point of departure into the FL* perspective: investigating the weak-coupling instabilities of the FL* will lead to good agreement with the current experimental observations. 
We shall conclude with an outlook in the final section.

\section{Review of some recent experiments}
\label{expts}
In the past few years, a number of remarkable experiments have given both direct and indirect evidence for the appearance of charge-order in the underdoped cuprates. Charge (and spin)-stripes have been known to exist in the Lanthanum-based cuprates \cite{tranquada,SKRMP03} for a long time now. In these materials, close to a hole-doping of $p=1/8$, the spin and charge stripes are known to have the most pronounced signatures; this is also where the superconducting $T_c$ goes all the way to zero. However, in this review we shall focus only on the non-La-based cuprates,  where charge-order is not accompanied by any spin-order. There are important differences 
between the properties of these two `classes' of cuprates. 

A preliminary indication of charge-order in Bi$_2$Sr$_2$CaCu$_2$O$_{8+\delta}$ came from STM experiments, which imaged the periodic modulation in the density of states near the vortex cores in the Meissner state \cite{JH02} and in the high-temperature pseudogap state \cite{AY04}. In YBa$_2$Cu$_3$O$_y$, NMR measurements have been able to detect field-induced long-range static charge order, predominantly on the oxygen sites, without any sign of spin-order \cite{MHJ11, MHJ13}. Ultrasound measurements at high-fields \cite{DLB13} have further corroborated a transition to a long-range charge ordered state.  The most clear signatures of any charge density wave correlations should come from X-ray. Indeed, incommensurate charge order has now been detected with resonant \cite{Ghiringhelli12,DGH12} and hard \cite{Chang12} X-ray scattering, both at zero and high fields. The wave vector has consistently been found to be directed along the copper-oxygen bonds and decreases with increasing hole-doping \cite{Hayden13}, a trend that is the exact opposite of what has been known to be the case in the La-based-cuprates \cite{MV09}. Moreover, there are strong indications that the charge modulation is highest on the bonds connecting the Cu sites \cite{Kohsaka07, DGH13}. A conclusion that is safe to draw from all of these experiments is that there exist incommensurate charge density wave correlations in the CuO$_2$ plane, which become static and (reasonably) long-ranged upon the application of a high magnetic field, and in general compete with superconductivity. 

\begin{figure}[h]
\begin{center}
\includegraphics[width=5in]{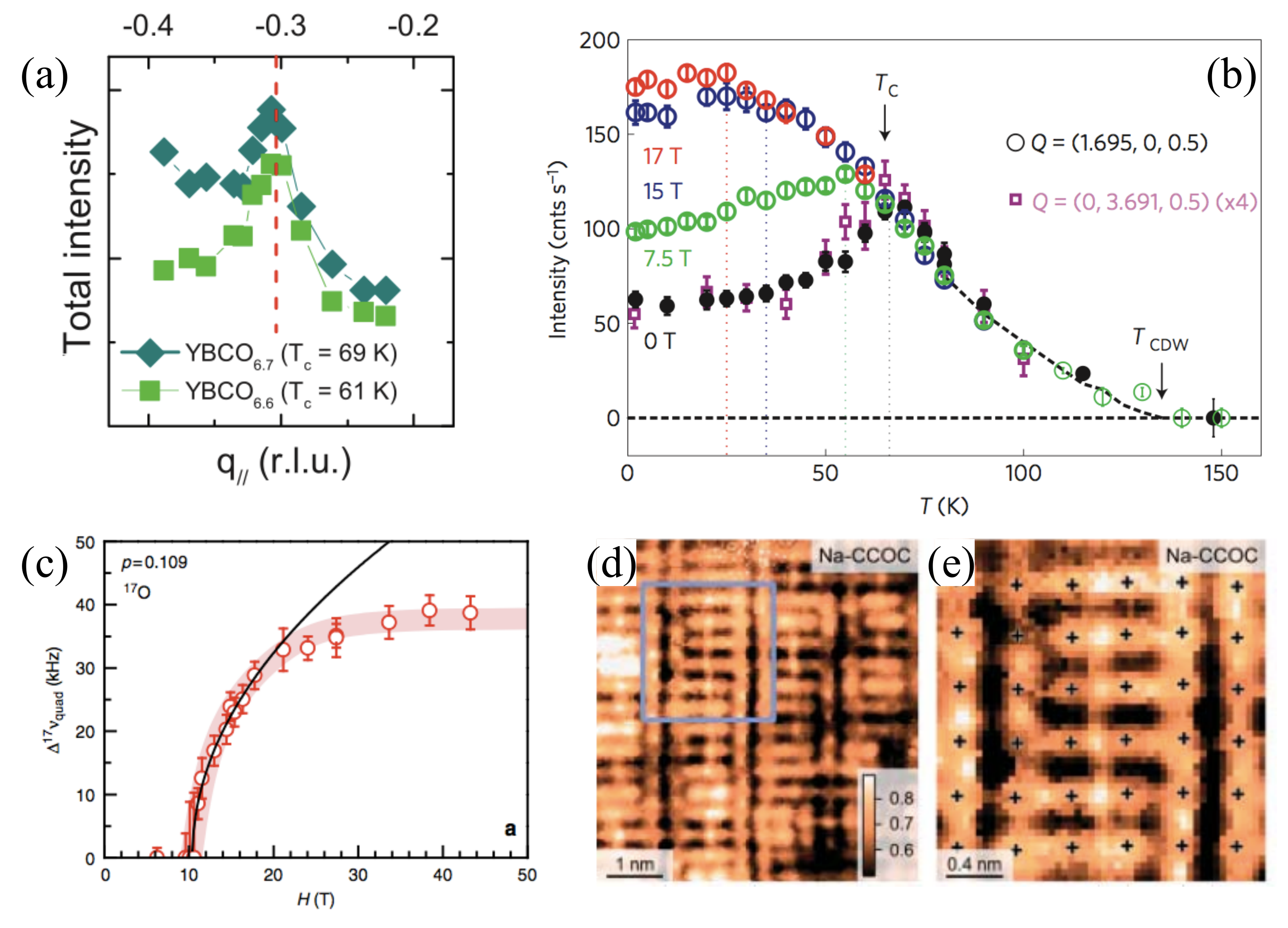}
\end{center}
\caption{(a) Dependence of the CDW intensity at $15$ K, showing a finite correlation length (as inferred from the width of the peak). Adapted from Ref.~\citenum{Ghiringhelli12}. (b) Temperature dependence of the peak intensity as a function of magnetic field. The onset shows a gradual non-mean-field like behavior. At zero-field, the intensity decreases below the superconducting $T_c$, showing competition between the two order-parameters. Adapted from Ref.~\citenum{Chang12}. (c) Quadrupole part of the splitting of the O line showing a field-induced transition. Adapted from Ref.~\citenum{MHJ13}. (d), (e) R-Maps taken at $150$ mV. Adapted from 
Ref.~\citenum{Kohsaka07}.}
\label{data}
\end{figure}

Some of the key experimental data that have shed light on the nature of the charge-order in the underdoped cuprates, is  reproduced in Fig.~\ref{data}. Fig. \ref{data}a from Ref.~\citenum{Ghiringhelli12} shows a peak in the total intensity at an incommensurate wavevector $\approx2\pi(0.3,0)$, in the superconducting phase. 
In order to study the interplay between charge-order and superconductivity, Chang {\it et al.\/} \cite{Chang12}, studied the onset of scattering intensity as a function of temperature both at zero and finite fields (Fig.\ref{data}b). They observed a gradual onset of charge density wave correlations at $T_{\tn{CDW}}$ (where $T_c<T_{\tn{CDW}}<T^*$; $T_{\tn{CDW}}$ typically correlates with $T^{**}$, as introduced earlier), with a subsequent decrease in the intensity below $T_c$ at zero-field. The feature for $T\leq T_c$ highlights the competition between the two order-parameters, which goes away when superconductivity is suppressed by a reasonably large magnetic field. The field-induced transition in the charge-order has also been investigated using NQR experiments\cite{MHJ13} (Fig.\ref{data}c). At moderate fields, such as those used in the X-ray and NQR experiments, the charge-order likely develops in the large `halos' surrounding the superconducting vortices{\footnote{In the extreme type-II limit ($\kappa=\lambda/\xi_{\tn{SC}}\rightarrow\infty$; $\lambda$ being the London penetration depth and $\xi_{\tn{SC}}$ the superconducting coherence length), the vortices really correspond to `superfluid' vortices \cite{EDSS01}.}} in the mixed-state. The field-induced transition occurs when the halos start to overlap, `locking' the orientation of the charge density wave over long distances. 

Finally, STM measurements show that the primary modulation resides on the Cu-O bonds and has a non-trivial pattern in real space 
(Fig.~\ref{data}d,e). The exact symmetry properties associated with this modulation have only been unveiled recently \cite{Fujita14}, as we shall discuss shortly.  

Let us first prescribe a systematic approach for expressing the charge-modulation on a two-dimensional square lattice. The charge density wave, or more appropriately {\it bond-density wave} (BDW), can be expressed as a bond-observable defined on sites $i,j$,
\beq
P_{ij}=\langle c_{i\alpha}^\dagger c_{j\alpha}\rangle = \sum_\Q \bigg[ \sum_\k P_\Q(\k) ~e^{i\k\cdot(\r_i-\r_j)} \bigg] e^{i\Q\cdot(\r_i+\r_j)/2}, \label{Pij1}
\eeq
where $c_{i\alpha}^\dagger$ creates an electron at site $i$ with spin $\alpha(=\uparrow, \downarrow)$. The action of various symmetry operations become quite transparent for the above parametrization in momentum space in terms of a relative momentum, $\k$, and a center-of-mass momentum $\Q$, as emphasized in Ref.~\citenum{AASScomment}. In momentum space, we can write 
Eq.~(\ref{Pij1}) as,
\beq
\left\langle c_{\k - \Q/2,\alpha}^\dagger c_{\k + \Q/2, \alpha}^{\vphantom \dagger} \right\rangle = P_\Q (\k) . \label{PQ}
\eeq
In contrast, the early work of Nayak \cite{CN00}, and numerous analyses since \cite{SCCN01,SCHYK08,vojta4,dunghai,dhlee,jhu}, have used the parameterization
\beq
\left\langle c_{\k ,\alpha}^\dagger c_{\k + \Q, \alpha}^{\vphantom \dagger} \right\rangle = F_\Q (\k) . \label{FQ}
\eeq
Of course, we can easily go back and forth between (\ref{PQ}) and (\ref{FQ}). However when we write $P_\Q (\k)$ as a periodic 
function of $\k$ alone, then the equivalent $F_\Q (\k)$ also depends upon $\Q$. In particular the state described by 
$P_\Q (\k) \sim \cos (k_x) - \cos (k_y)$, proposed in Refs.~\citenum{SSRLP13,MMSS10b}, is a $d$-form factor bond density wave which preserves time-reversal and which will play an important experimental role shortly. In contrast, the state $F_\Q (\k) \sim \cos (k_x) - \cos (k_y)$, considered by others, is a different state; for general $\Q$, it is a mixture of
components that are both even and odd under time-reversal, and so it is not a useful starting point for a symmetry analysis.

Conventional charge density waves with $P_\Q(\k)$ independent of $\k$ lead only to an on-site charge modulation, with the overall modulation period set by $2\pi/|\Q|$. However, this is not the case in the context of the cuprates. The experiments on at least two different families of the cuprates (BSSCO and Na-CCOC), which involves phase-sensitive STM \cite{Fujita14} and X-ray \cite{Comin14sym} measurements, have now unveiled the form factor $P_\Q(\k)$ to be predominantly of a $d$-wave nature, i.e. $P_\Q(\k) \sim (\cos k_x - \cos k_y)$. In addition, as already mentioned above, almost all the experiments point towards a strong evidence for the wavevector $\Q$ to be along the Cu-O bonds, i.e. $(\pm Q_0,0)$ and $(0,\pm Q_0)$, where $Q_0$ decreases with increasing hole-doping (and is therefore, incommensurate). For an underlying `large' putative Fermi surface, which is absent in this regime (as known from photoemission experiments, which only see Fermi-`arcs'), these wavevectors would nest regions in the vicinity of, but away from the antinodes: $(\pi,0)$ and $(0,\pi)$. 

On the other hand, there is a recent X-ray observation \cite{DGH14} in the La-based cuprates, which seems to indicate a dominant $s'-$form factor (where $P_\Q(\k) \sim (\cos k_x + \cos k_y)$), and this has been ascribed to the presence of magnetic 
stripe order in these compounds \cite{ATSS14,AK14}. The nature of the bond-density waves seen in the two different `classes' of cuprate superconductors is therefore qualitatively different---both in terms of the predominant form-factor, and, in terms of the doping dependence of the wavevector. 

\begin{figure}[h]
\begin{center}
\includegraphics[width=4.5in]{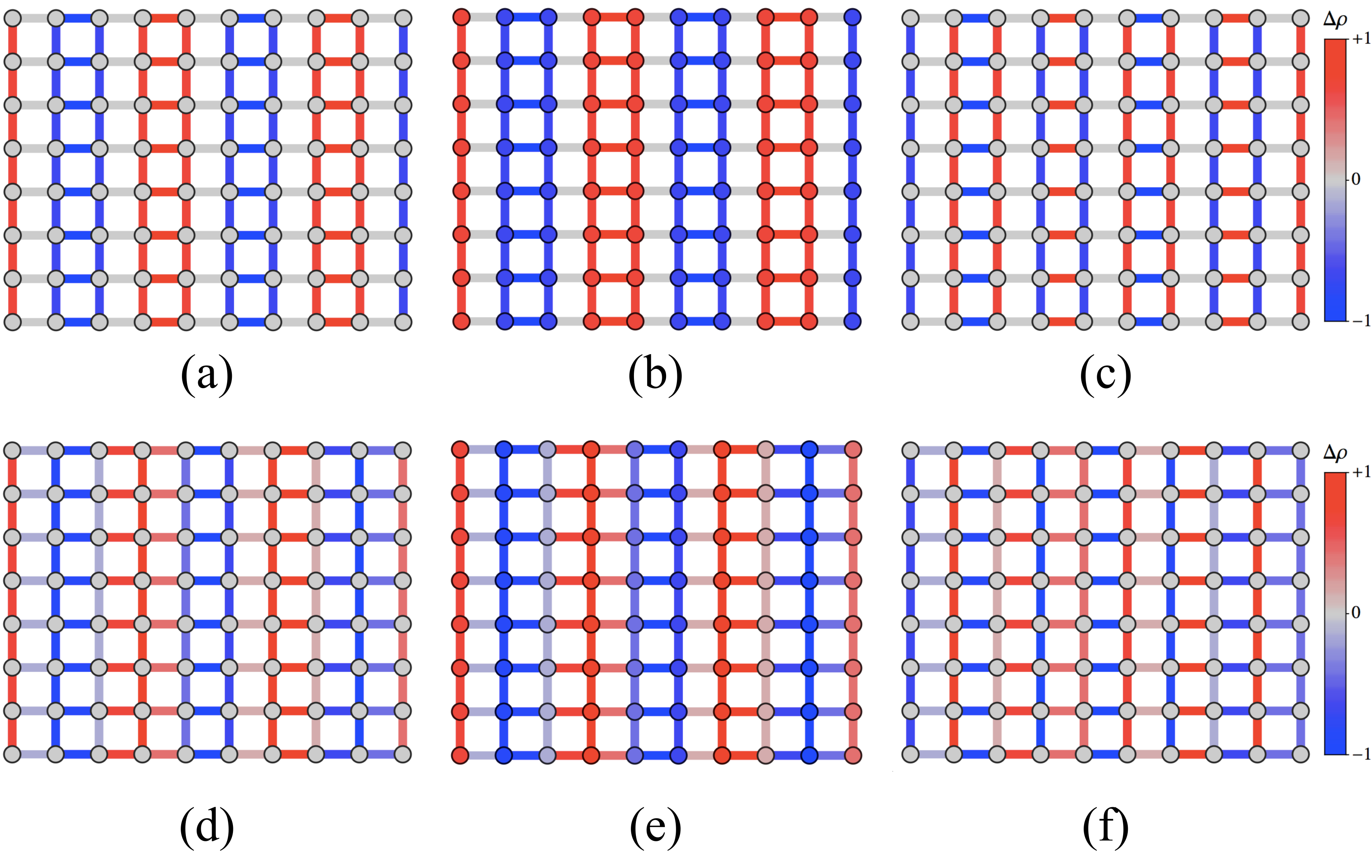}
\end{center}
\caption{Real-space visualization of unidirectional BDW (charge-stripe) with components: $P_s(\k) = P_s, P_{s'}(\k) = P_{s'}(\cos k_x + \cos k_y)$, and $P_d(\k) = P_d(\cos k_x - \cos k_y)$. (a)-(c) we plot the charge modulation for
a commensurate wavevector, $\Q = 2\pi(\frac{1}{4},0)$, while (d)-(f) plot the same quantity for an incommensurate
wavevector, $\Q = 2\pi(0.3,0)$. The parameters used are: (a), (d) $P_s = 0, P_{s'} = 1, P_d = 0$. (b), (e) $P_s =1, P_{s'} =1, P_d =0$. (c),(f) $P_s =0, P_{s'} =0, P_d =1$. We have included phases in the definitions of $P_{s,s',d}$ in order to make the charge distribution “bond-centered” for the case of the commensurate wavevector; however, other choices of the phases are also allowed. Adapted from Ref.~\citenum{DCSS14b}.}
\label{bdw}
\end{figure}

In Fig.~\ref{bdw} we provide an illustration of unidirectional BDWs with different components of $P_\Q(\k)$ in real space for both commensurate as well as incommensurate wavevectors. It is not a coincidence that the patterns of Fig.\ref{bdw} (c), (f) and the data of Kohsaka et al. \cite{Kohsaka07} in Fig.\ref{data}(d), (e) look remarkably similar. This comparison has recently been carried out to a remarkable degree of precision in the phase-sensitive work of Fujita et al.\cite{Fujita14}, which has pinned down the symmetry to be predominantly $d$-wave.

\begin{figure}[h]
\begin{center}
\includegraphics[width=5in]{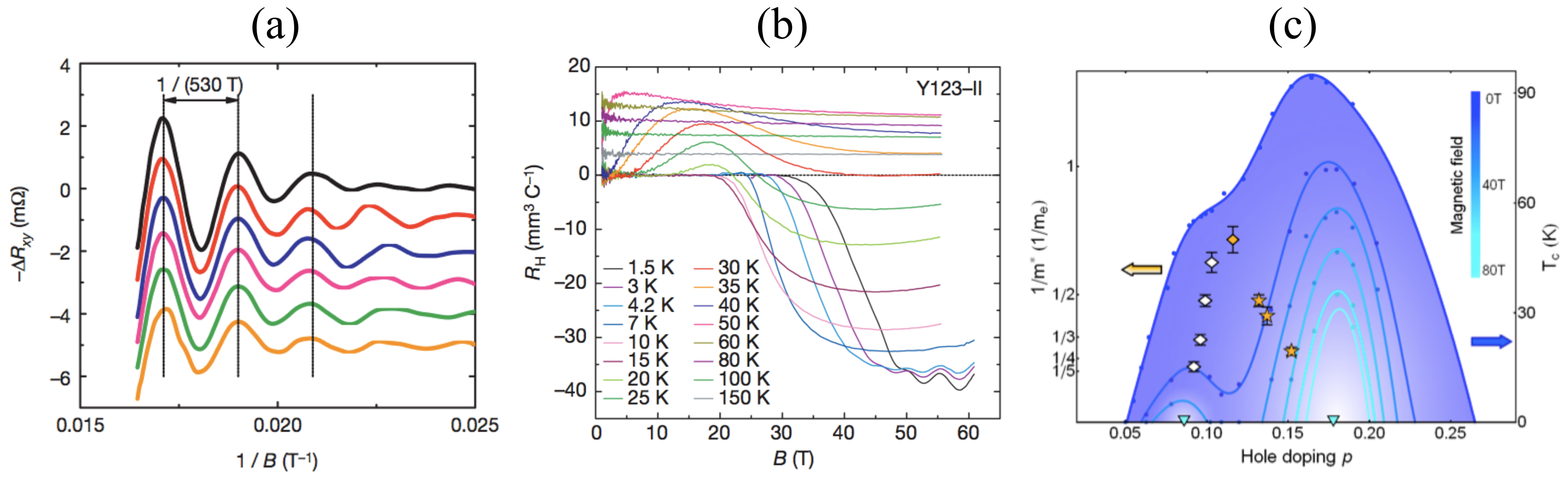}
\end{center}
\caption{(a)  The oscillatory part of the Hall resistance (after subtracting the monotonic background at $T=2$ K), as a function of inverse magnetic field, $1/B$. Adapted from Ref.~\citenum{LT07}. (b) The Hall coefficient as a function of $B$ at different temperatures. Adapted from Ref.~\citenum{DLB07}. (c) The blue curves represent $T_c$, at different $B$ and are supposed to highlight the connection between effective mass enhancement and the strength of superconductivity. Adapted from Ref.~\citenum{BJR14}.}
\label{dataqo}
\end{figure}

A key question now is whether the presence of the charge density wave order may also explain the quantum oscillations observed at low temperatures and high magnetic fields \cite{LT07,SSNHGL12}. We shall discuss this topic at some length in the next section, but it is worthwhile to also review some of the most striking experimental aspects here. The first observation of quantum-oscillations \cite{LT07} provided clear signatures of a {\it reconstructed} pocket with an area approximately $2\%$ of the Brillouin zone (corresponding to a frequency $\sim 530$T; see Fig.\ref{dataqo}a). This is fundamentally different from the observations in extremely overdoped cuprates, where quantum-oscillations observed a `large' Fermi surface \cite{Proust08}. The key question then was to figure out the density-wave order responsible for giving rise to the reconstruction in the underdoped cuprates. An interesting fact that was realized soon after the initial discovery of oscillations, is related to the nature of carriers in the system. In spite of doping holes into the system, transport-properties were most consistent with the presence of primarily electron-like quasiparticles, as evidenced by the {\it negative} Hall coefficient \cite{DLB07} (Fig.\ref{dataqo}b). This prompted the search for a small {\it electron-pocket}, that would explain both of these observations. 

More recently, by extracting the effective cyclotron mass of the electron-like quasiparticles{\footnote{By studying the temperature dependence of the oscillation amplitudes and fitting to the Lifshitz-Kosevich form.}} and by noting the critical dopings where they appear to diverge (Fig.\ref{dataqo}c), the location of the underlying quantum critical points, corresponding to the onset of charge-order, have been determined \cite{SSNHGL12, BJR14}. These QCPs also coincide with the dopings around which the $T_c(B)$-domes, as deduced from the resistive transition in a field ($B$), are centered about. We direct the readers to some recent review articles for further details regarding the quantum oscillation experiments \cite{SSNHGL12, MHJCP13}.

\section{Phenomenology at lower temperatures}
\label{pheno}

After the above review of some of the key experiments, let us start addressing the questions that we had raised at the end of the introduction. Here we shall present semi-phenomenological models of the ordering instabilities at low temperatures.

\subsection{Onset of charge-order and superconductivity}
\label{sec:o6}

As mentioned in the previous section, a peculiar feature associated with charge-order in the underdoped cuprates, is the highly {\it non mean-field-like} nature of the onset, at a temperature that is typically lower than the pseudogap temperature, $T^*$. In fact this behavior is reminiscent of the onset of antiferromagnetism in the parent insulating antiferromagnetic state of La$_2$CuO$_4$ \cite{Keimer92}. We now understand this gradual onset of intensity{\footnote{The finite transition temperature arises solely due to the non-zero interlayer coupling.}} in the structure factor over a broad range of temperatures in terms of the {\it angular fluctuations} of an O(3) field (corresponding to the N$\acute{\tn{e}}$el order) in two spatial dimensions \cite{CHN89}. 

The above observation then begets the question if the onset of charge-order can be understood in terms of the angular-fluctuations of an order-parameter \cite{KBE13}, and if so, can one write down an effective field theory that describes these fluctuations. Let us first note the symmetries associated with the above problem: O(2)$\times$O(2)$\times$O(2)$\times \mathbb{Z}_2$. These correspond to, charge-conservation, $x-$translations, $y-$translations and the $x\leftrightarrow y$ symmetries. In addition, we use the fact that the system also preserves time-reversal and inversion symmetries. 

\begin{figure}
\begin{center}
\includegraphics[width=4in]{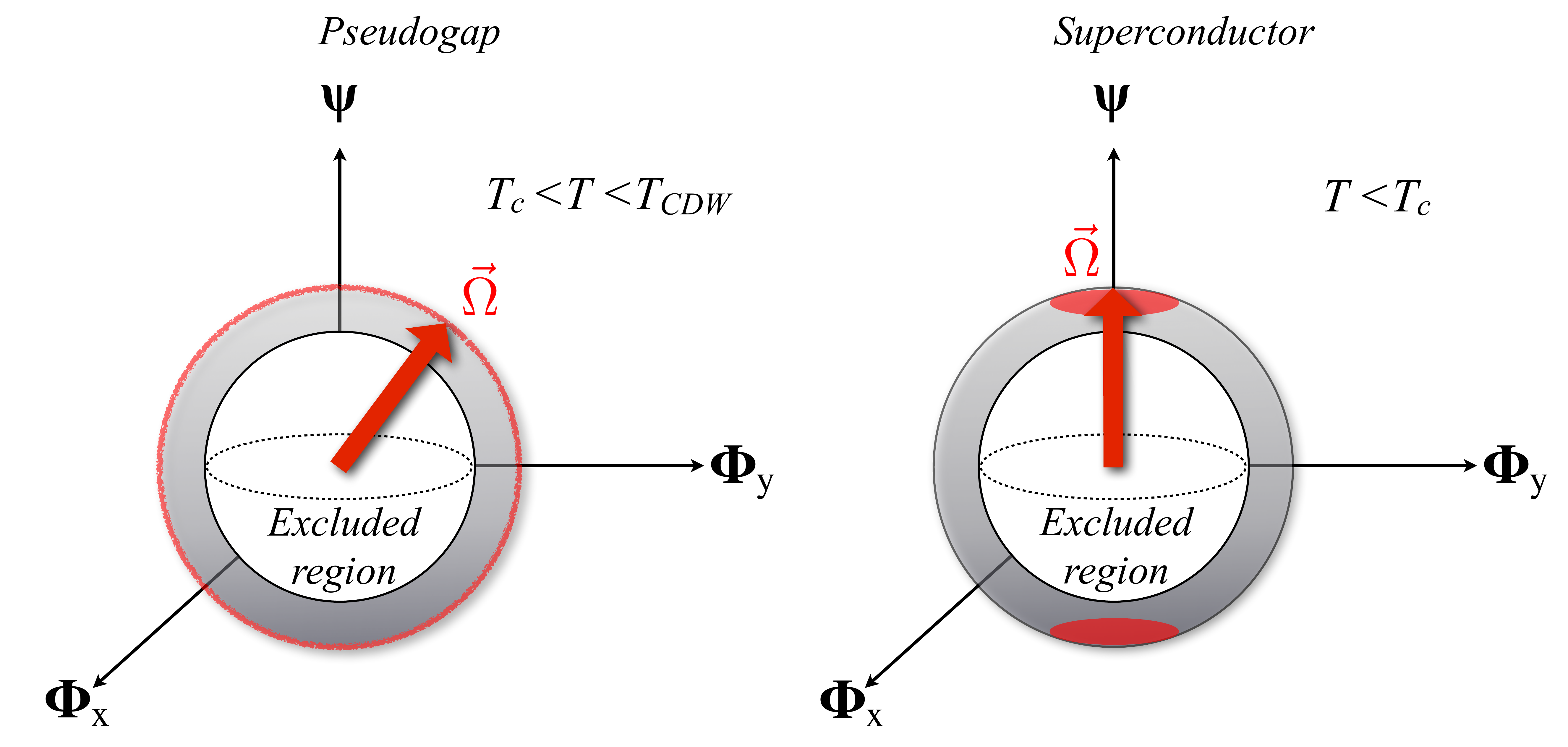}
\end{center}
\caption{The angular fluctuations of the six-component order-parameter, $\vec{\Omega}$, describe the non-mean field onset of charge density wave correlations below $T_{\tn{CDW}}$. Due to a finite $g>0$, $\vec{\Omega}$ fluctuates only along the superconducting directions for $T<T_c$, the Kosterlitz-Thouless transition temperature corresponding to the $O(2)$ component associated with $\Psi$. For $w<0$, the $Z_2$ invariance associated with $x\leftrightarrow y$ can be broken spontanteously.}
\label{nlsm}
\end{figure}
 
The onset can then be best understood in terms of the angular fluctuations associated with a six-component order parameter \cite{LHSS14}: 
\beq
\vec{\Omega}=(\tn{Re}\Psi,\tn{Im}\Psi,\tn{Re}\Phi_x,\tn{Im}\Phi_x,\tn{Re}\Phi_y,\tn{Re}\Phi_y), 
\eeq
where $\Psi$ represents the superconducting component and $\Phi_{x,y}$ represent the incommensurate charge density wave order parameters along $x$ and $y$ directions{\footnote{Note that the region in the vicinity of $\vec{\Omega}=0$ is forbidden, i.e. the amplitude fluctuations of $|\vec{\Omega}|$ are completely forbidden below $T_{\tn{CDW}}$ and probably onset at an even higher temperature.}}. The fluctuations are assumed to be of  a purely classical and thermal nature and are postulated to be described by the following non-linear $\sigma$-model (NL$\sigma$M),
\beq
Z&=&\int{\cal{D}}\vec{\Omega}(\vec r)\delta(\vec{\Omega}^2-1)~\tn{exp}\bigg(-\frac{\cal{S}_\tn{cl}}{T}\bigg),\\
{\cal{S}}_\tn{cl}&=&\frac{\rho_{\tn{S}}}{2}\int d^2r\bigg[|\nabla\Psi|^2+\lambda(|\nabla\Phi_x|^2 + |\nabla\Phi_y|^2)\nonumber\\
&&+g(|\Phi_x|^2 + |\Phi_y|^2)+w(|\Phi_x|^4 + |\Phi_y|^4)\bigg].
\label{o6}
\eeq
In the above, $\rho_{\tn{S}}$ and $\rho_{\tn{S}}\lambda$ control the helicity moduli of the superconducting and of the density wave order respectively. The coupling $g$ breaks the symmetry explicitly between the $\Psi$ and $\Phi_x, \Phi_y$ directions. It sets the relative energetic cost of ordering between the superconducting and density wave directions. We take $g > 0$, so that superconductivity is preferred at low temperatures (figure \ref{nlsm}). The coupling $w$ imposes the square lattice point group symmetry on the density wave order.  

Note that within this framework, in the intermediate temperature range $(T<T_{\tn{CDW}})$ over which the charge order correlations have been observed, we can choose not to include the effect of antiferromagnetic correlations explicitly; they can be absorbed into the phenomenological parameters of the above classical model. However, for the onset of the pseudogap at $T^*$, these can't be ignored.

\begin{figure}
\begin{center}
\includegraphics[width=5in]{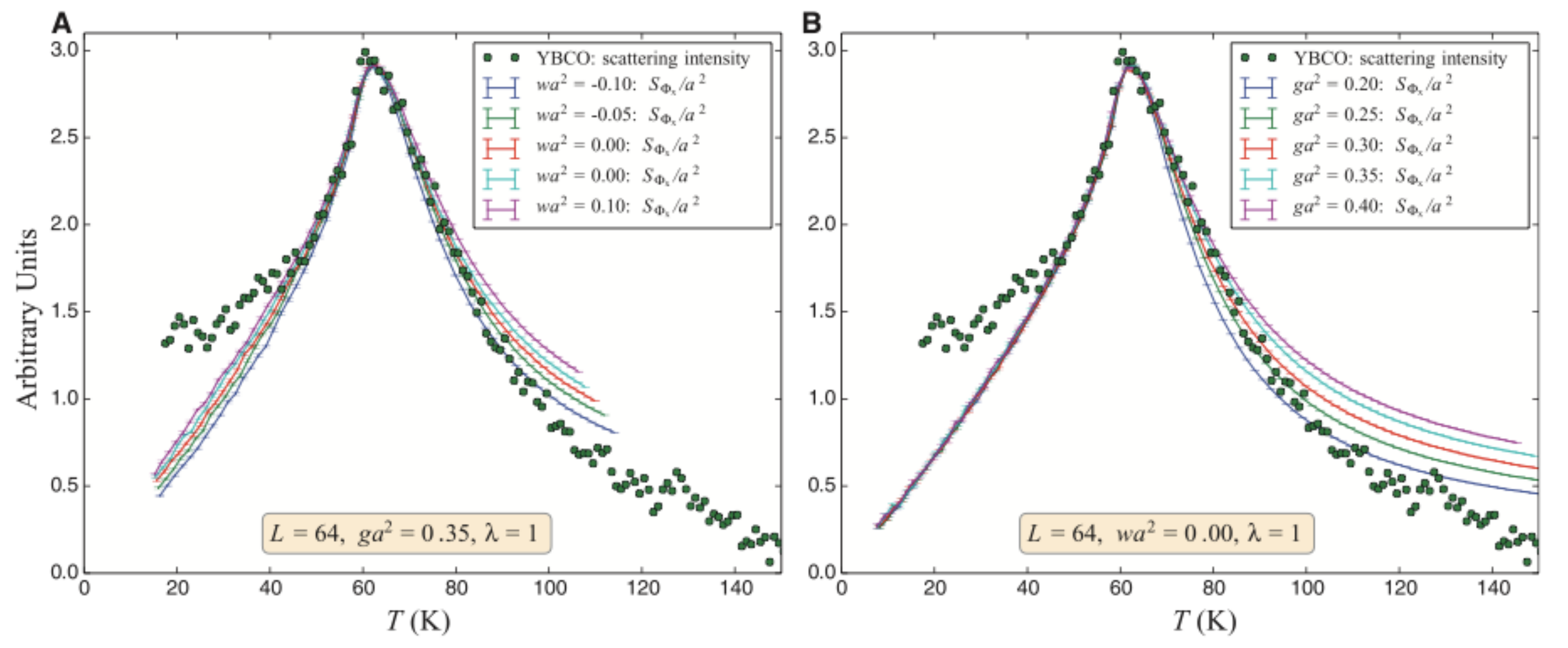}
\end{center}
\caption{Comparison of X-ray data for the CDW scattering intensity to the structure-factor calculated in Monte Carlo simulations of ${\cal{Z}}$ for different parameter sets. Adapted from Ref.\citenum{LHSS14}.}
\label{sf}
\end{figure}

It is possible to add higher-order terms to the above action, without significant qualitative differences in the results. The above action was studied using (classical) Monte-Carlo simulations and within a `$1/N$' expansion (where $N$ corresponds to the number of components of $\vec{\Omega}$; $N=6$ for the theory in Eq.\ref{o6}) \cite{LHSS14}. The results for the structure factor and comparison to experiments on YBCO are shown in Fig. \ref{sf}. The agreement with experimental data is very good, except at low and high temperatures. In the absence of impurities, that would otherwise pin the charge-order and give rise to a finite intensity at low temperatures, the structure-factor goes to zero. On the other hand, at higher temperatures, the onset of amplitude fluctuations, which are ignored in the above setup, would also decrease the intensity. 

The theory in Eq.~(\ref{o6}) has also been applied to measurements of the diamagnetic susceptibility \cite{NPO10,JC13} in YBCO. This measures the superconducting
$\Psi$ component of the O(6) order parameter, complementing the X-ray measurements of the charge order represented by $\Phi_x$, $\Phi_y$.
It was found \cite{LHSS14b} that the same set of parameters can provide a reasonable simultaneous fit to both the X-ray and diamagnetism measurements. 
Thus the O(6) theory provides a surprising quantitative connection between two very different classes of experiments, and this supports
its validity as a theory of fluctuating orders in the low temperature pseudogap regime.

\subsection{Quantum oscillations}

Some of the most interesting aspects of the underdoped cuprates involve the nature of the fermionic excitations. However, in the above description so far, we have ignored the Fermionic degrees of freedom altogether. In order to study these questions, here we couple the underlying electrons to $d$-wave superconducting and $d$-form-factor charge density wave order parameters and explore the fermionic excitations at low temperatures and high magnetic fields.
As discussed in Section~\ref{expts}, this is the regime where quantum oscillations have been observed. It is not entirely clear what the upper critical field, $H_{c2}$, is for the underdoped cuprates and if the fields, $B\sim 40$T, where oscillations are first observed are already higher than $H_{c2}$. Moreover, whether the application of large fields completely suppresses the pseudogap is also a topic of ongoing debate. These are questions that need to be resolved in order to realize {\it what} gets reconstructed to give rise to the small electron-pocket.

It is nevertheless a reasonable starting point to study the problem of reconstruction on top of a `large' Fermi surface in the presence of a (sufficiently) long-ranged BDW{\footnote{Such that the cyclotron-radius associated with an applied magnetic field is smaller than the correlation length of the BDW.}}.

\begin{figure}
\begin{center}
\includegraphics[width=4in]{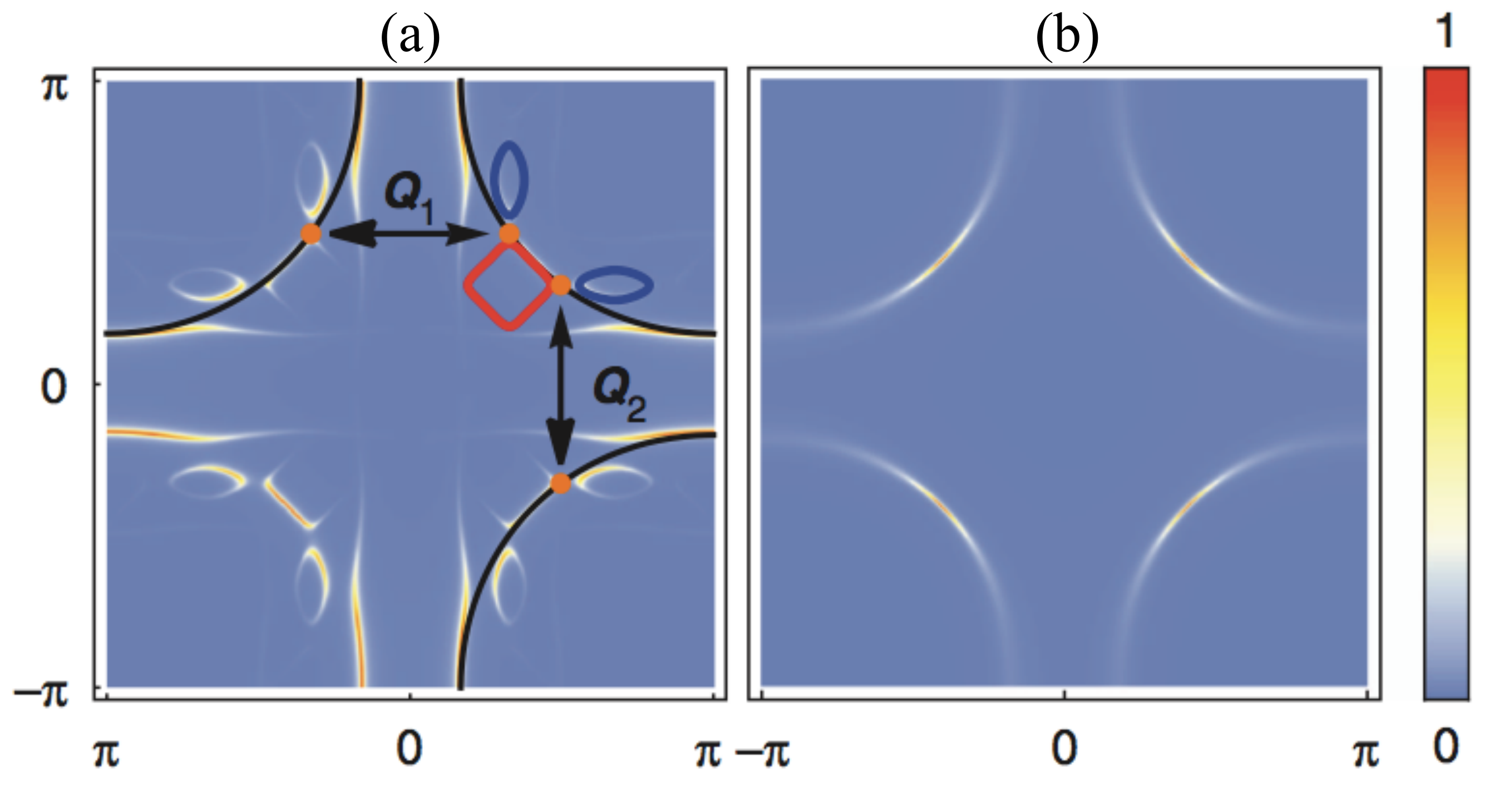}
\end{center}
\caption{(a) The electron spectral function in the presence of long-range bidirectional BDW (with wavevectors $\Q_1$ and $\Q_2$) at zero magnetic field (in the unfolded Brillouin zone). In black, the Fermi surface used for our computation. The BDW causes reconstruction of the Fermi surface and the formation of an electron-like pocket (red) and two hole-like pockets (blue). (b) Electron spectral function in the presence of fluctuating superconducting and BDW correlations. Adapted from Ref.~\citenum{AADCSS14}.}
\label{qope}
\end{figure}

The spectral function, in the unfolded Brillouin zone, for a large Fermi surface in the presence of long-range{\footnote{Such long-ranged BDW is likely to be present only in a strong magnetic field and will not be seen in photoemission experiments.}} BDW  is shown in Fig.\ref{qope}a. A small {\it nodal} electron-pocket \cite{HS11} can be seen (shown in red) \cite{AADCSS14}, which is likely responsible for giving rise to the oscillations corresponding to $\sim 530$T. A significant prediction of our computation is that an invariable consequence of reconstructing a large Fermi surface is the presence of small {\it hole-}pockets closer to the antinodes (shown in blue). There have been recent reports for the existence of these hole-pockets\cite{LThole14} at frequencies close to those predicted by our analysis.\cite{AADCSS14}

We have also extended the above analysis of reconstruction of the large Fermi surface to the case of fluctuating order: this would apply at zero field
in the same lower temperature regime where we described X-ray and diamagnetism experiments in Section~\ref{sec:o6}.
We describe the fluctuating
order by the O(6) model, and compute the fermion self energy at the one loop level. Typical results appear
in Fig.~\ref{qope}b, showing `Fermi arc' spectra in the nodal regions. In photoemission experiments, such Fermi arc spectra are also seen at 
significantly higher temperatures, where the O(6) model of fluctuating order cannot be reasonably applied. We believe a rather different model
is necessary at higher temperatures, and we will turn to this in Section.~\ref{flstar}.

Recent experiments have also detected a divergence in the effective cyclotron mass, $m^*$, as extracted from quantum-oscillations, near optimal doping \cite{BJR14}. It has been argued \cite{TS14} that the critical like divergence arises from the angle-dependent Fermi-velocity, $v_F(\theta)$, along the nodal electron-pocket, and is in fact dominated by the corners of the pocket{\footnote{Since $m^*=\frac{1}{2\pi}\int\frac{d\theta}{v_F(\theta)}$.}}. However, signatures of this divergence are absent in any measurement performed at zero field, that is dominated by the `light' nodal quasiparticles, such as transport.    

Despite the agreement between the computation of the large Fermi surface reconstruction by the BDW and the quantum oscillation 
experiments, there are a number of outstanding puzzles that remain. Notice that in addition to the closed pockets in Fig.\ref{qope}(a), there are a number of {\it open Fermi-sheets}, that are bound to be present as long as the BDW potential is small. Even though these don't contribute to oscillations, the large density of states along these sheets would contribute to various thermodynamic measurements. However, recent measurements of the Knight-shift \cite{KS10} and specific heat \cite{GB11} in fields of upto $\sim 45$ T do not seem to be consistent with the availability of such large density of states at the Fermi-level. One could argue that if the BDW potential were to be arbitrarily large, it could gap out these portions of the Fermi surface as well. However, in the presence of such strong {\it incommensurate} density-wave order, quantum oscillations are likely to be disrupted completely \cite{AMSK14}.

The above thermodynamics experiments seem to suggest that whatever is causing the pseudogap to open up close to the antinodes remains unaffected at low to intermediate fields. This leaves us with two options: ({\it i\/}) The reconstruction occurs on top of a `pseudogapped' Fermi surface, so that it is really the arc that gets reconstructed. We shall revisit this scenario further in Section~\ref{flstarinst}, or,   ({\it ii\/}) The reconstruction happens in a `vortex-liquid' phase, where the vortices associated with $d$-wave superconductivity are fluctuating \cite{MR13}. However, within either of these two scenarios, the nodal electron-pocket is likely to survive, since this region of the Fermi surface remains unaffected, both by $d$-wave superconductivity and by the pseudogap.

Another issue that remains to be resolved with regard to the nature of the charge-order responsible for the oscillations at high fields is as follows. Experiments at zero-fields seem to suggest the presence of a stripe-like order, while the nodal electron-pocket scenario requires a nearly checkerboard order. In order to relate these two regimes, there could be two possibilities:  ({\it i\/}) There is a field induced transition from a stripe-like BDW at zero, or, low-fields to a nearly checkerboard-like BDW at high-fields, or, ({\it ii\/}) The system actually consists of differently oriented domains with stripe-like order, either within each CuO$_2$ plane, or within different planes (but with a strong interlayer tunneling) \cite{AVMPHSR14}, where the typical domain sizes, $\xi_{\tn{domain}}$, are smaller than the cyclotron-radius. Future experiments will hopefully be able to shed light on this issue.

\section{Fractionalized Fermi liquid at higher temperatures}
\label{flstar}

We now turn to the higher temperature regime just below $T^\ast$, significantly above temperatures at which there is any indication of
appreciable fluctuating order, and so the analysis of the previous section cannot directly apply.
Here, as we have discussed earlier, the primary experimental indications of the pseudogap are the suppression in the spin susceptibility, and the appearance of `Fermi arc' spectra in the electron spectral function. 

\begin{figure}
\begin{center}
\includegraphics[width=2.3in]{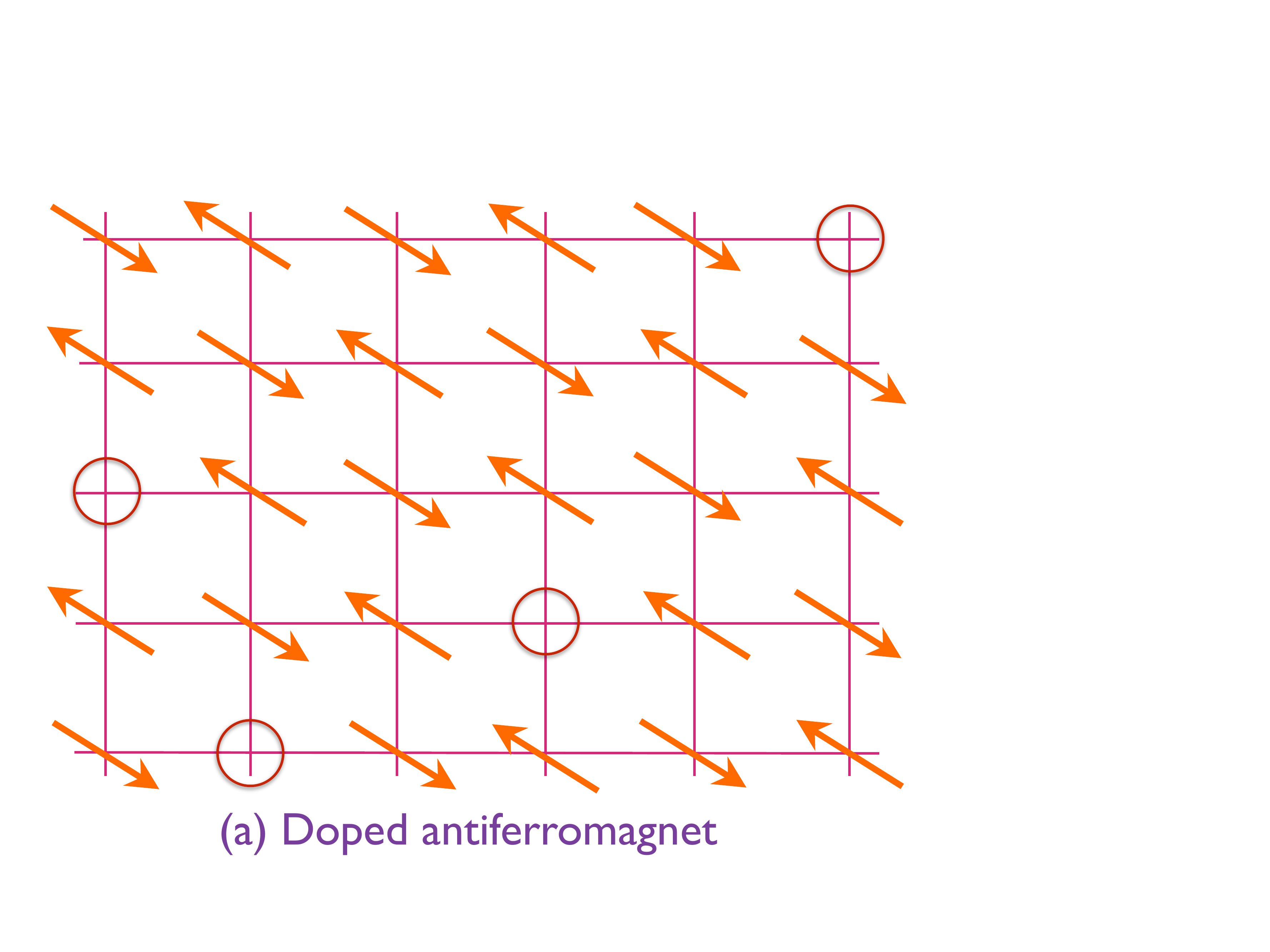}~~~~
\includegraphics[width=2.3in]{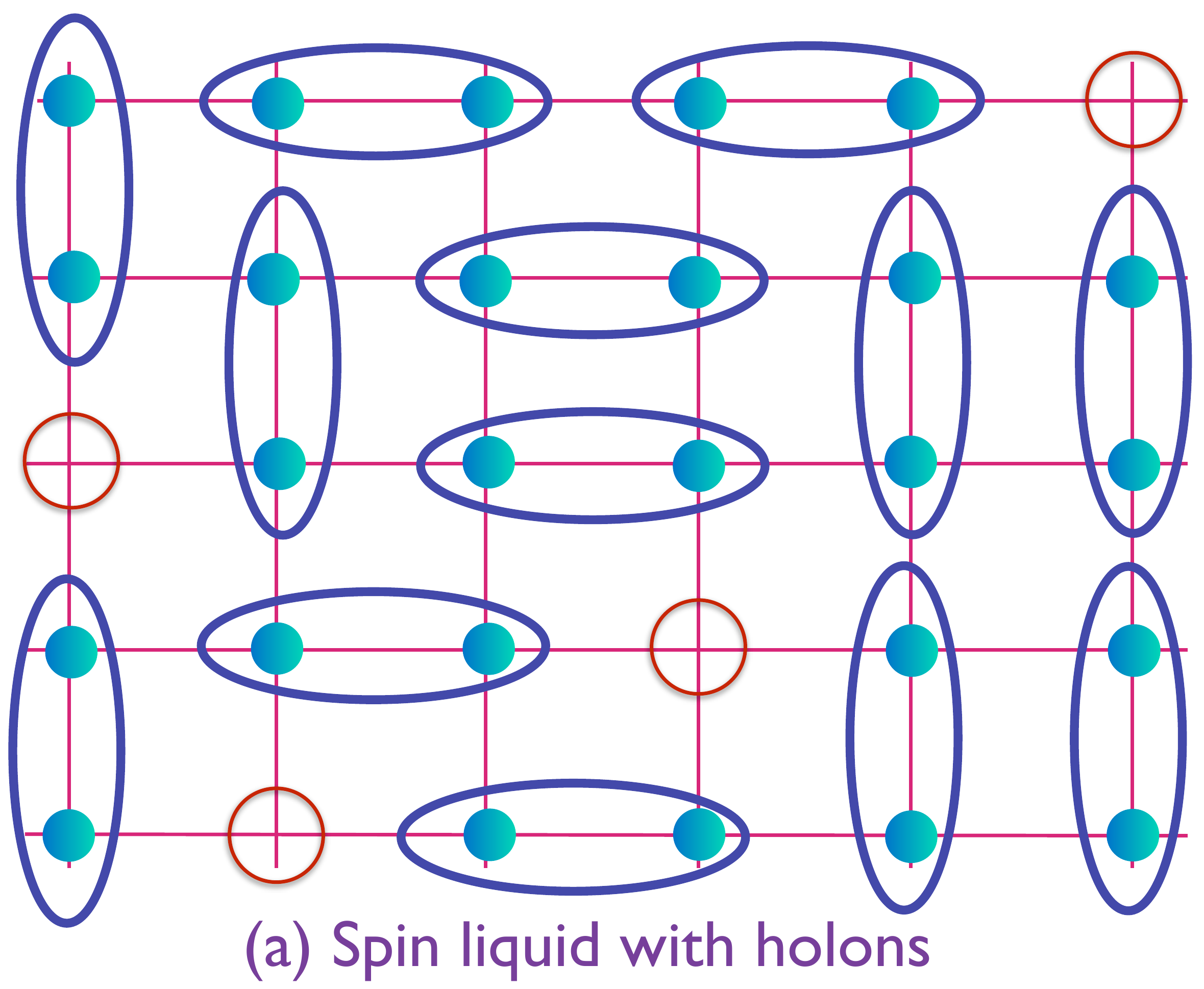}
\end{center}
\caption{(a) A density of $p$ holes per unit cell in an ordered antiferromagnet. The holes are the open circles, while the arrows denote
the orientation of the electron spins on the remaining sites. Note that there is a density of $1+p$ holes with respect to the band insulator
in which there are 2 electrons on each site. (b) Spin liquid with a density $p$ of spinless holons of charge $+e$. The ellipses represent spin-singlet
pairs of electrons in the state $\left( \left| \uparrow \downarrow \right\rangle - \left| \downarrow \uparrow \right\rangle \right)/\sqrt{2}$. }
\label{dimer12}
\end{figure}

\begin{figure}
\begin{center}
\includegraphics[width=2.3in]{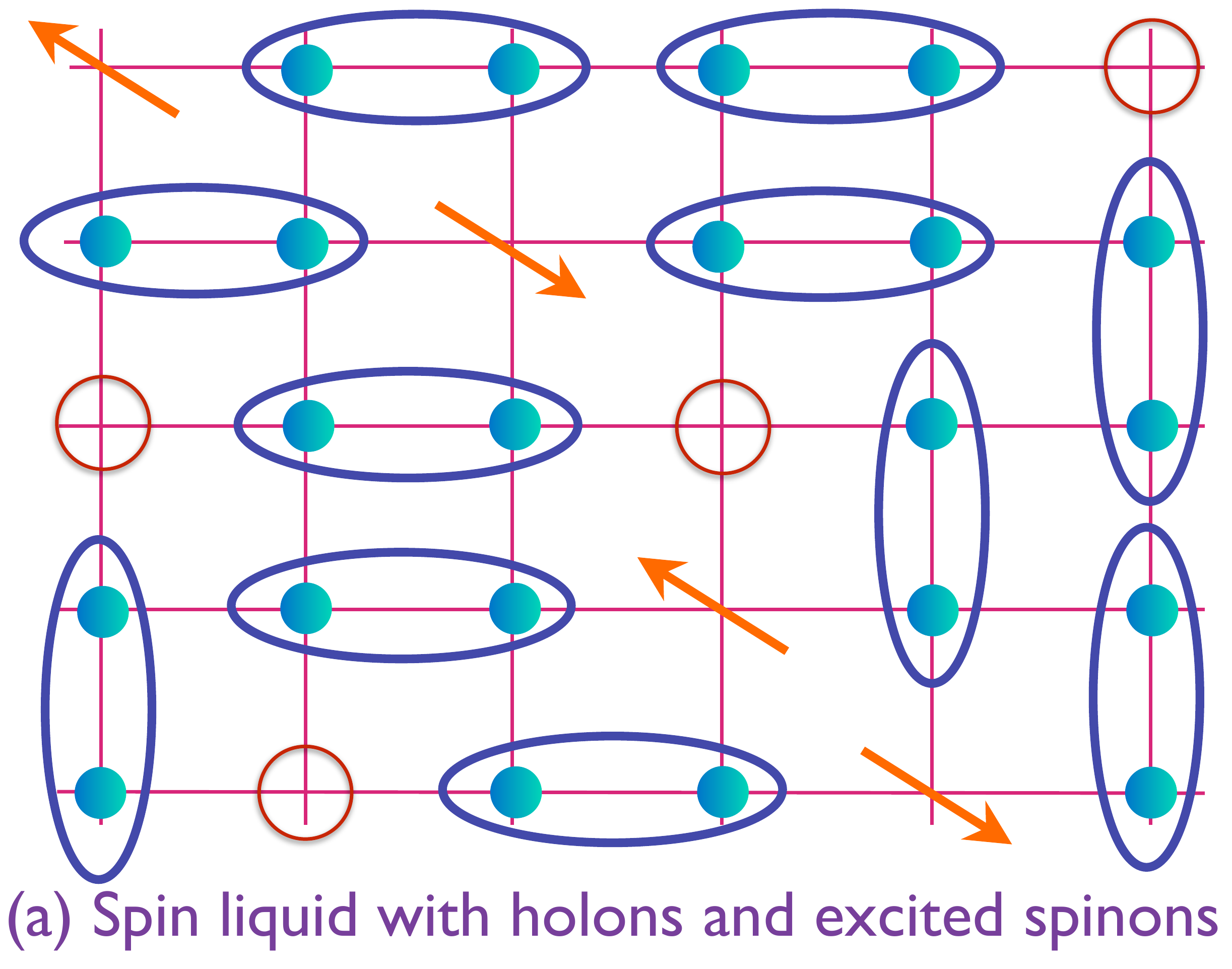}~~~~
\includegraphics[width=2.3in]{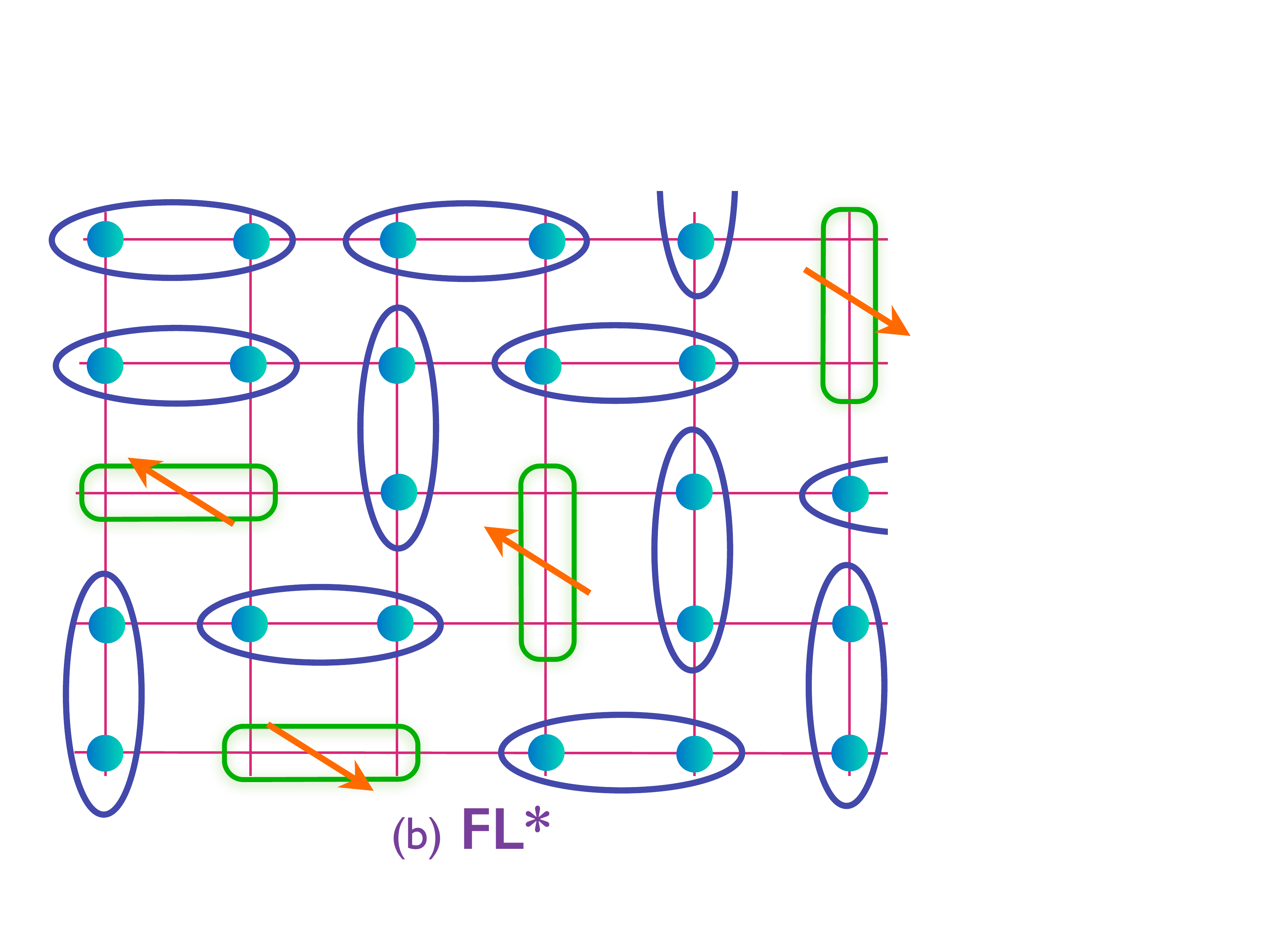}
\end{center}
\caption{(a) An excited state of the doped spin liquid, with neutral $S=1/2$ spinon excitations represented by the arrows.
Both the holons and the spinons can move independently in the spin liquid background. Note the holon and spinon on nearest neighbors in the 
middle of the figure: the electron can easily exchange positions between these sites, and this can lead to a holon-spinon bound state
in which the electron is in a bonding orbital between the two sites. (b) The FL* state: All holons have paid the energy cost to create a spinon from the spin liquid, and the resulting bound state is represented by the green dimers. The green dimers are fermions carrying charge $+e$ and spin $S=1/2$, and their motion in the spin liquid background realizes a metal with Fermi surfaces enclosing an area equivalent to a density of $p$ fermions of spin $S=1/2$.}
\label{dimer34}
\end{figure}

We review a model of this pseudogap metal as a `fractionalized Fermi liquid' (FL*). While related states have been derived by many different theoretical
methods \cite{Wen96,YRZ06,RK07,YQSS10}, we begin with a simple toy model description \cite{MPAASS15} which generalizes the quantum dimer models \cite{SKDRJS87,DRSK88} of insulating spin liquids. (We believe that recent results from dynamical mean field theory \cite{Ferrero09,Gull13} also point to a low
energy effective theory of the pseudogap in terms of such a dimer model \cite{MPAASS15}).
Consider a doped antiferromagnet with $p$ holes per unit cell, as shown in Fig.~\ref{dimer12}a. 
Note that there are $p$ holes with respect to the
antiferromagnet, but the number of holes with respect to the filled band insulator is $1+p$. As $p$ increases, we imagine that the antiferromagnetic
order is quickly destroyed, and we obtain a state with $p$ holes in a spin liquid background as shown in Fig.~\ref{dimer12}b. Now all the spins 
of the antiferromagnet have been paired into singlets, and these singlets can resonate with each other \cite{pwa87}. The holes in this spin liquid background 
carry no spin and charge $+e$ as they move around: so these are `holons'. This doped spin liquid state also has neutral $S=1/2$ excitations known
as `spinons', as shown in Fig.~\ref{dimer34}a.

So far we have just described the well-known structure of a doped spin liquid. To obtain the FL*, we need one further step: suppose there is an attractive
potential between the holons and spinons so that they form bound states with charge $+e$ and spin $S=1/2$ (a rationale for this attractive
potential is given in the caption of Fig.~\ref{dimer34}a). 
For simplicity, let us 
take the strong coupling limit
in which this bound state involves only nearest neighbor sites, and so can be represented by a dimer, as shown{\footnote{
The bound state is represented in Fig.~\ref{dimer34}b by an arrow, and this represents an electron which is in a bonding orbital
between the two sites. In a three-band model, this can be associated with an electron on the O orbital between the Cu orbitals.
However, our analysis does not require the three-band model, and applies equally to the one-band Hubbard model.}}
 in Fig.~\ref{dimer34}b.
If the binding energy is large enough, every holon will pay the energy cost needed to create a spinon out of the spin liquid background,
and we obtain a spin liquid doped only by fermionic dimers carrying charge $+e$ and spin $S=1/2$.

The dynamics of this FL* is now described by a quantum dimer model \cite{MPAASS15} in which the fermionic dimers and the original spin singlet pairs resonate with each other. This leads to motion of the fermionic dimers, and as there is a dilute gas of them, we can expect them to form a metallic state
with a Fermi surface. The quasiparticles near the Fermi surface will now have the same quantum numbers as in a Fermi liquid, but they differ in key aspects:
\begin{itemize}
\item The volume enclosed by the Fermi surface in a FL* is equivalent to a density of $p$ spin $S=1/2$ particles of charge $+e$.
In contrast, in a Landau Fermi liquid the equivalent density would be $1+p$.
\item The FL* has a second independent sector of low energy `topological' excitations. In the quantum dimer model,
these are associated with the resonances of the spin-singlet dimers, which are represented in analytical formulations by emergent gauge fields. 
\item The quasiparticles near the Fermi surface are neutral under the emergent gauge field. In the quantum dimer model presented here,
this is a consequence of the fact that the fermions are themselves also dimers. Indeed, the fermions couple to the gauge field non-minimally
as {\it dipoles\/}. This coupling is not strong enough to disrupt the quasiparticle nature of the Fermi surface excitations.
\end{itemize}
The dimer model picture demonstrates the inevitability of the above characteristics of the FL*: the fermionic dimers have density $p$, while the remaining
sites are required to have spin-singlet dimers representing the gauge field. In Ref.~\citenum{MO00,TSMVSS04}, general arguments have been given
which demonstrate that any violation by a Fermi surface of the $1+p$ Luttinger volume requires the presence of low energy topological excitations.

The  following subsections will move beyond this dimer toy-model picture of the FL* phase, and review field theoretic analyses which
yield a metallic state with the same basic characteristics. Such analyses allow us to compute the Fermi surface structure, understand potential low $T$ instabilities of the FL* metal,
and also connect to other phases in the global cuprate phase diagram. 

\subsection{Field-theoretic description}

We begin our description of the FL* metal by first reviewing a standard description for the evolution of 
antiferromagnetism in metallic phases of the one-band Hubbard model. This model is widely believed to contain the essential physics of the cuprates \cite{pwa87}, for electrons hopping on the sites of a square lattice,
\beq
H_\tn{hubbard}&=&H_t+H_U,\\
H_t&=&-\sum_{i<j}t_{ij} c_{i\alpha}^\dagger c_{j\alpha} - \mu\sum_i c_{i\alpha}^\dagger c_{i\alpha},\\
H_U&=&U\sum_i\bigg(n_{i\uparrow}-\frac{1}{2} \bigg)\bigg(n_{i\downarrow}-\frac{1}{2} \bigg),
\eeq
where $t_{ij}$ represent the hopping parameters, $U$ is the on-site Coulomb repulsion, $\mu$ represents the chemical potential and $\alpha=\uparrow,\downarrow$ are the spin-indices. The Fermions satisfy the standard anti-commutation relations, $\{ c_{i\alpha}, c_{j\beta}^\dagger\}=\delta_{ij}\delta_{\alpha\beta}$ and $\{c_{i\alpha},c_{j\beta}\}=0$. This completely defines the problem at hand, that has eluded an exact solution. 

Anticipating the metallic state to be in the vicinity of an antiferromagnetic instability close to half-filling, we use the exact operator equation,
\beq
U \bigg(n_{i\uparrow}-\frac{1}{2} \bigg)\bigg(n_{i\downarrow}-\frac{1}{2} \bigg) = -\frac{2U}{3} \vec{S}_i^2 + \frac{U}{4},
\eeq
valid on each site, $i$, and where $\vec{S}_i=c_{i\alpha}^\dagger \vec\sigma_{\alpha\beta} c_{i\beta}/2$. Upon decoupling the interaction via a Hubbard-Stratanovich transformation, we obtain
\beq
\tn{exp}\bigg(\frac{2U}{3}\sum_i\int d\tau \vec{S}_i^2 \bigg) = \int {\cal{D}}\vec{J}_i(\tau) \tn{exp}\bigg(-\sum_i\int d\tau \bigg[ \frac{3}{8U} \vec{J}_i^2 -\vec{J}_i\cdot\vec{S}_i\bigg] \bigg).
\eeq
We can now integrate out the fermions from the action and look for the saddle point of the resulting action for $\vec{J}_i$. This leads to the N$\acute{\tn{e}}$el state with a wavevector, $\K=(\pi,\pi)$. In the long-wavelength limit, it is then useful to introduce a field, $\vec\vp_i$,
\beq
\vec{J}_i = \vec\varphi_i e^{i\K\cdot\r_i}.
\eeq
This is then the familiar route towards arriving at the ``spin-fermion" model,
\beq
{\cal{Z}}&=&\int {\cal{D}}c_\alpha {\cal{D}}\vec\vp ~\tn{exp}(-{\cal{S}}),\\
{\cal{S}}&=&\int d\tau \bigg[\sum_\k c_{\k\alpha}^\dagger ( \partial_\tau - \ve_\k) c_{\k\alpha} - \lambda \sum_i c_{i\alpha}^\dagger \vec\vp_i\cdot\vec\sigma_{\alpha\beta} c_{i\beta} e^{i\K\cdot\r_i}\bigg] \nonumber\\
&+&\int d\tau~d^2\r \bigg[ \frac{\rm{v}^2}{2} (\nabla_\r\vec\vp)^2 + \frac{1}{2} (\partial_\tau\vec\vp)^2 + \frac{s}{2} \vec\vp^2 +\frac{u}{4} (\vec\vp^2)^2\bigg].
\label{SF}
\eeq
\begin{figure}[h]
\begin{center}
\includegraphics[width=4in]{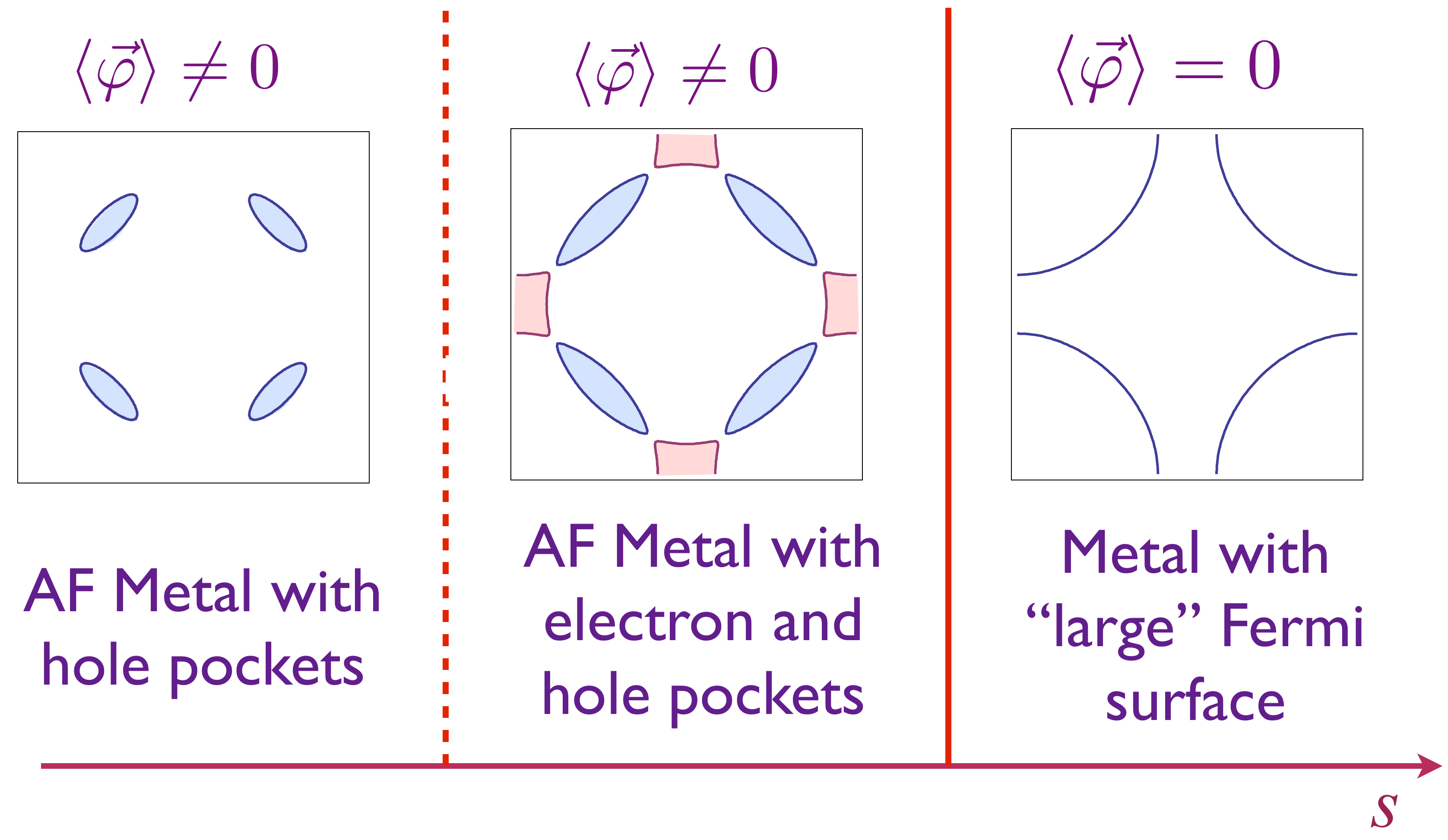}
\end{center}
\caption{The evolution from a `large' Fermi surface (for $\langle\vec\vp\rangle=0, ~s>s_c$) to a reconstructed Fermi surface with electron (red) and hole (blue) pockets in the presence of antiferromagnetic (AF) order with a finite $\langle\vec\vp\rangle\neq0$ ($s<s_c$) across a SDW QCP at $s=s_c$. By moving sufficiently deep inside the SDW phase, one has a metallic state with only hole-pockets in the presence of a large $\langle\vec\vp\rangle$.}
\label{sdw}
\end{figure}
The first term in ${\cal{S}}$ is just the kinetic energy of the fermions, and the second term is the ``Yukawa" coupling term that leads to the spin-fluctuation mediated scattering of fermions from one portion of the Fermi surface to the other. The momentum of the boson, $\vec\vp$, is small,
and $\rm{v}$ represents a characteristic spin-wave velocity.
 The last line in the above action is just the $\vp^4-$ field theory for an $N=3$ order-parameter. 

At mean-field level, as we tune the value of $s$, we go from a metallic Fermi-liquid phase with a `large' Fermi surface ($s>s_c$) to a metal with reconstructed electron and hole pockets ($s<s_c$). The gap due to the SDW order parameter opens up at the `hot-spots', where $\ve_{\k}=\ve_{\k+\K}=0$. With a decreasing $s$, as $\langle \vec\vp\rangle$ increases in magnitude, only the hole pockets remain (Fig.\ref{sdw}).  

\begin{figure}[h]
\begin{center}
\includegraphics[width=4.3in]{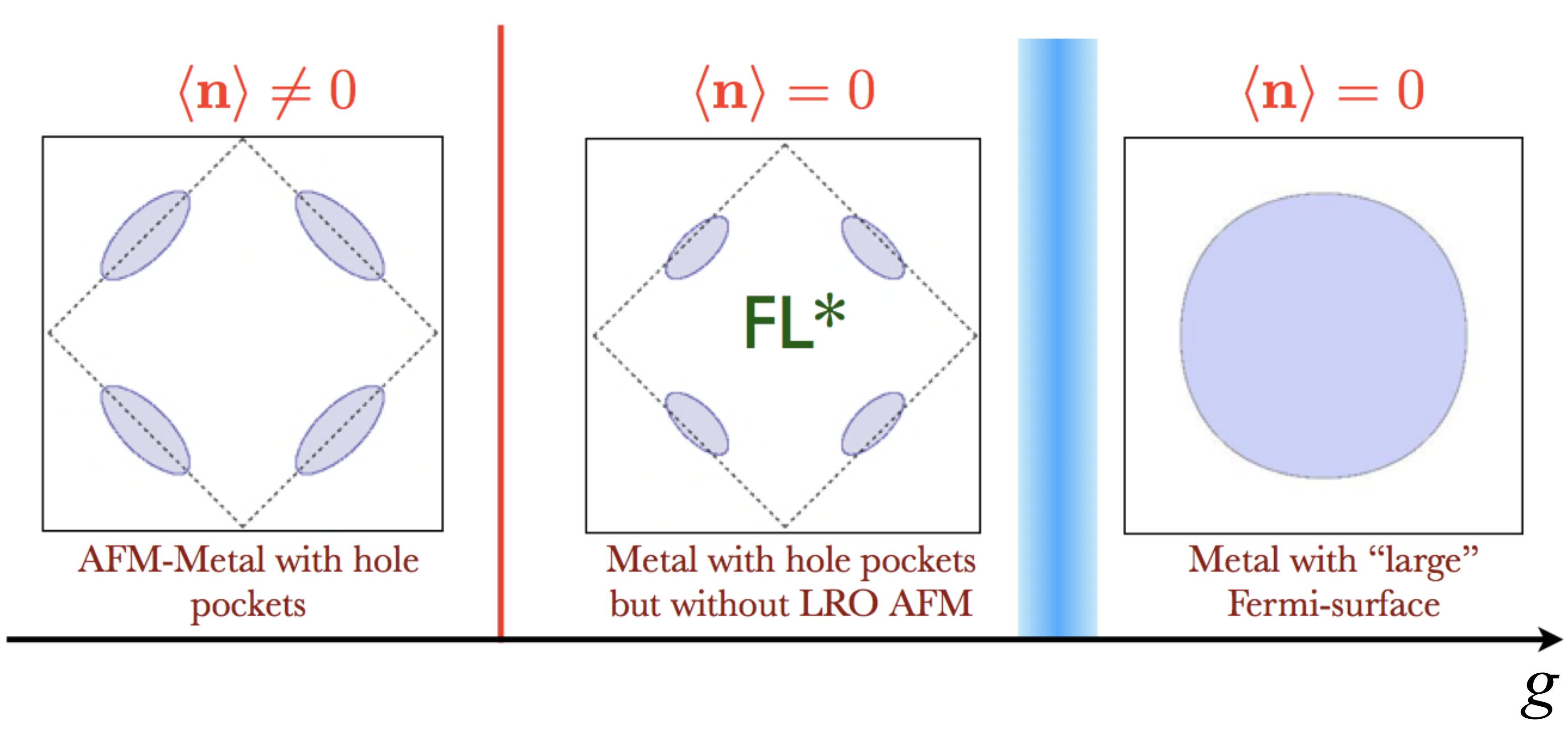}
\end{center}
\caption{Alternative route to Fig.~\ref{sdw} for the loss of antiferromagnetic order in a metal in a theory with {\it angular\/} fluctuations
of the fixed-length order $\n$. The middle phase is FL*. The evolution from FL* to the large Fermi surface Fermi liquid
on the right is addressed in Ref.~\citenum{DCSS14c}.}
\label{loss}
\end{figure}

So far, apart from possible exotic behavior at quantum critical points, we have only obtained conventional Fermi liquid phases.
To obtain the FL*,  we have to switch from the above description of SDW order by a ``soft-spin'' $\vec\vp$ field with large amplitude fluctuations,
to a ``hard-spin'' perspective in which we have primarily {\it angular\/} fluctuations of the antiferromagnetic order. So we replace
$\vec\vp$ by a unit vector field $\n$,
\beq
\vec\vp ~ \Rightarrow \n \quad, \quad \n^2 = 1.
\label{fixedlength}
\eeq
The key utility of such a formulation in terms of the unit-length $\n$ field is that it allows us to consider a route to
``quantum-disordering'' the SDW order in which topological ``hedgehog'' tunneling events are suppressed \cite{MV04}; in contrast, the 
perturbative analysis of the soft-spin theory necessarily proliferates hedgehogs at the zeros of the $\vec\vp$ field.
In such a hard-spin theory, we argue that the evolution of phases in Fig.~\ref{sdw} can be replaced by the more exotic route shown in 
Fig.~\ref{loss}.
Now there is an intermediate phase in which there are hole pockets enclosing a volume equivalent to fermions with density $p$, but without
antiferromagnetic order: this is the FL* phase. 

\begin{figure}[h]
\begin{center}
\includegraphics[width=5.2in]{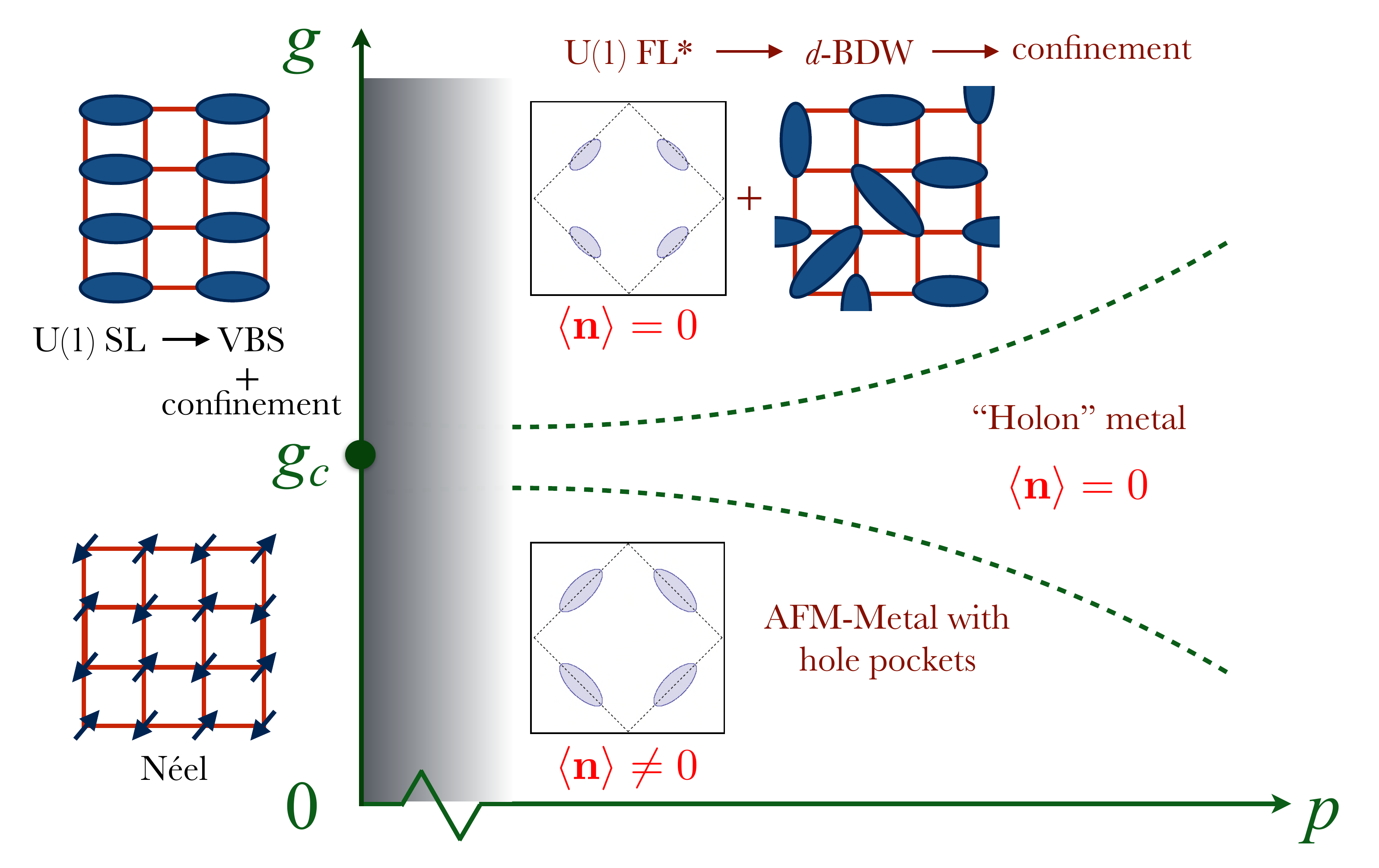}
\end{center}
\caption{Adapted from Ref.~\citenum{DCSS14b}.
The vertical axis represents some frustrating coupling in an insulating antiferromagnet which can drive it into a valence bond solid (VBS)
across a deconfined quantum critical point at $g=g_c$.
}
\label{dcp}
\end{figure}

Let us turn to an explicit presentation of the theory of a metal with angular fluctuations of antiferromagnetic order.
We are only interested in the long-wavelength fluctuations of $\n_i$ while retaining the full lattice dispersions for the fermions. 
Applying (\ref{fixedlength}) to (\ref{SF}) we obtain
the imaginary time Lagrangian, $\L=\L_f+\L_n+\L_{fn}$, with
\beq
\L_f&=&\sum_{i,j}c_{i\alpha}^\dagger\bigg[(\partial_\tau-\mu)\delta_{ij} - t_{ij}\bigg]c_{j\alpha} +\tn{h.c.},\\ 
\L_n&=&\frac{1}{2g}\int d^2\r ~[(\partial_\tau\n)^2+\rm{v}^2(\nabla\n)^2],\\
\L_{fn}&=&\lambda\sum_i e^{i\K\cdot\r_i}~ \n_i\cdot c_{i\alpha}^\dagger \vec\sigma_{\alpha\beta} c_{i\beta}.
\eeq
Now $s$ has been replaced by $g$ to tune the strength of quantum fluctuations associated with the AFM order parameter. 
We assume $g$ is a generic coupling measuring the degree of frustration in the insulating antiferromagnet, which can drive the insulator
into a non-magnetic state with valence bond solid (VBS) order across a deconfined quantum critical point \cite{senthil1}.
It is therefore useful to discuss the 
phase diagram of a general class of frustrated doped antiferromagnets,
as a function of the coupling $g$ and the charge density $p$, as shown in Fig.~\ref{dcp}.

For $g<g_c$, the above model has long-range antiferromagnetic order, $\langle\n\rangle\neq0$ (with a correlation length, $\xi\rightarrow\infty$), which reconstructs the large Fermi surface. The interesting regime is when $g \geq g_c$, which we argue is relevant to the physics of the non-La-based cuprates.
The deconfined critical point $g=g_c$ and $p=0$ is expressed not in terms of the $\n$ fields, but in terms of the spinor $z_\alpha$, with $\n_i=z_{i\alpha}^*\vec\sigma_{\alpha\beta}z_{i\beta}$ and a U(1) gauge field $A_\mu$.
At the same time \cite{YQSS10}, one tranforms the underlying electrons, $c_{i\alpha}$, to a new set of spinless fermions, $\psi_{ip}$,
\beq
c_{i\alpha}&=&R_{\alpha p}^i \psi_{ip},~~~\tn{where}\\
R_{\alpha p}^i&=&\left( \begin{array}{cc}
z_{i\uparrow} & -z_{i\downarrow}^* \\
z_{i\downarrow} & z_{i\uparrow}^*
\end{array} \right), \label{cRz}
\eeq
is a spacetime dependent SU(2) matrix ($\sum_\alpha|z_{i\alpha}|^2=1$) and the fermions $\psi_{ip}$ carry opposite charges $p=\pm 1$ under the same emergent U(1) gauge transformation. Note that the above parametrization introduces a large SU(2) `gauge' redundancy and it is possible to embed the U(1) gauge-theory within the larger SU(2) gauge structure. This has been used to describe the QCP around optimal doping recently, in terms of a `Higgs'-transition without any broken symmetries \cite{SS09,DCSS14c} 

We begin our discussion of the phase-diagram in Fig.~\ref{dcp} by first considering the effect of non-zero $p$ at the critical coupling $g=g_c$. We are not interested here in the very small values of $p$ at which the $\psi_p$ fermions get  effectively localized \cite{RK07} and exclude this regime in the phase-diagram (shown as the grey-shaded region). At higher hole densities, the holons form a Fermi surface, and this can then quench the $A_\mu$ fluctuations via Landau damping \cite{RKMMSS08,TS08}; the holon Fermi surface is then stable\footnote{At low temperatures, the holons can pair to form a composite Boson that is neutral under the $A_\mu$ field, condensation of which leads to the holon superconductor \cite{RK08}.} against confinement \cite{MHTS04,SSL08} and we obtain the holon metal, also referred to as the U(1) {\it algebraic charge liquid}, shown in Fig.~\ref{dcp}.

Let us now turn to $g>g_c$. Now there is at least one additional length scale, the spin correlation length, $\xi$. This length should be compared with the spacing between
the holons $\sim 1/\sqrt{p}$. When $\xi \ll 1/\sqrt{p}$, we have to first consider the influence of a non-zero $\xi$ on the gauge theory of the deconfined critical point.
As described in Refs.~\citenum{senthil1,senthil2}, in this regime there is a crossover to a Coulomb phase in which the $A_\mu$ field mediates a logarithmic Coulomb force. 
This Coulomb force binds the $z_\alpha$ and $\psi_p$ quanta into gauge-neutral fermions \cite{RK07,RK08,YQSS10} (an additional attractive force is also provided by the ``Shraiman-Siggia term'' \cite{RK07,YQSS10,SS88}). At longer scales, these gauge-neutral fermions start to notice each
other via the Pauli principle, and so they form Fermi surfaces leading to the FL* state of interest here.

As discussed above, the emergent photon gives rise to binding of the $\psi_p$ fermions and the $z_\alpha$ spinons into gauge-neutral objects. However, there are two such combinations,
\beq
F_{i\alpha}\sim z_{i\alpha}\psi_{i+},~~G_{i\alpha}\sim \varepsilon_{\alpha\beta}z^*_{i\beta}\psi_{i-},
\eeq
where $\varepsilon_{\alpha\beta}$ is the unit antisymmetric tensor. Note that unlike the bound-state Fermionic dimers in the previous discussion of the toy-model, the above bound-states arise from spinons and holons on the same site. The physical electronic operator has a non-zero overlap with both of these combinations,
\beq
c_{i\alpha}\equiv Z(F_{i\alpha}+G_{i\alpha}),
\eeq
where $Z$ is a quasiparticle renormalization factor that is nonlocal over $\xi$. Over distances that are larger than $\xi$, where there is no net AFM order, the $F_\alpha$ and $G_\alpha$ fermions preferentially, but not exclusively, reside on the different sublattice sites.  

Based on symmetry considerations alone, we can write the following effective Hamiltonian \cite{YQSS10} for a FL* metal realized by
$F_{i\alpha}$ and $G_{i\alpha}$,
\beq
H_{\tn{eff}}=&-&\sum_{i,j} t_{ij}(F_{i\alpha}^\dagger F_{j\alpha} + G_{i\alpha}^\dagger G_{j\alpha}) + \lambda\sum_i e^{i\K\cdot\r_i}(F_{i\alpha}^\dagger F_{i\alpha} - G_{i\alpha}^\dagger G_{i\alpha}) \nonumber\\
&-& \sum_{i,j}\tilde{t}_{ij}(F_{i\alpha}^\dagger G_{j\alpha} + G_{i\alpha}^\dagger F_{j\alpha}).
\eeq
Once again, the $t_{ij}$ represent the hopping matrices corresponding to a large Fermi surface, $\lambda$ represents the potential due to the local AFM order (at distances shorter than $\xi$). The terms proportional to $\tilde{t}_{ij}$ represent the analogs of the ``Shraiman-Siggia" (SS) terms \cite{SS88} which couple the $F$ and $G$ particles. When $\tilde{t}_{ij}\neq0$, one obtains hole-like pockets centered away from $(\pi/2,\pi/2)$.

The Green's function for the electronic operator is given by \cite{YQSS10},
\beq
G^c(\k,\omega)&=&\frac{Z^2}{\omega-\xi^+_\k-\lambda^2/[\omega-\xi^-_{\k+\K}]},
\label{ACL}
\eeq
where the dispersions, $\xi_\k^+,~\xi_\k^-$ are,
\beq
\xi_\k^+&=&\ve_\k+\tilde\ve_\k,~~\xi_\k^-=\ve_\k-\tilde\ve_\k,\\
\ve_\k&=&-2t_1(\cos(k_x)+\cos(k_y))-4t_2\cos(k_x)\cos(k_y)\nonumber\\
&&-2t_3(\cos(2k_x)+\cos(2k_y))-\mu,\\
\tilde\ve_\k&=&-\tilde{t}_0-\tilde{t}_1(\cos(k_x)+\cos(k_y)).
\label{disp}
\eeq

\begin{figure}[h]
\begin{center}
\includegraphics[width=4.7in]{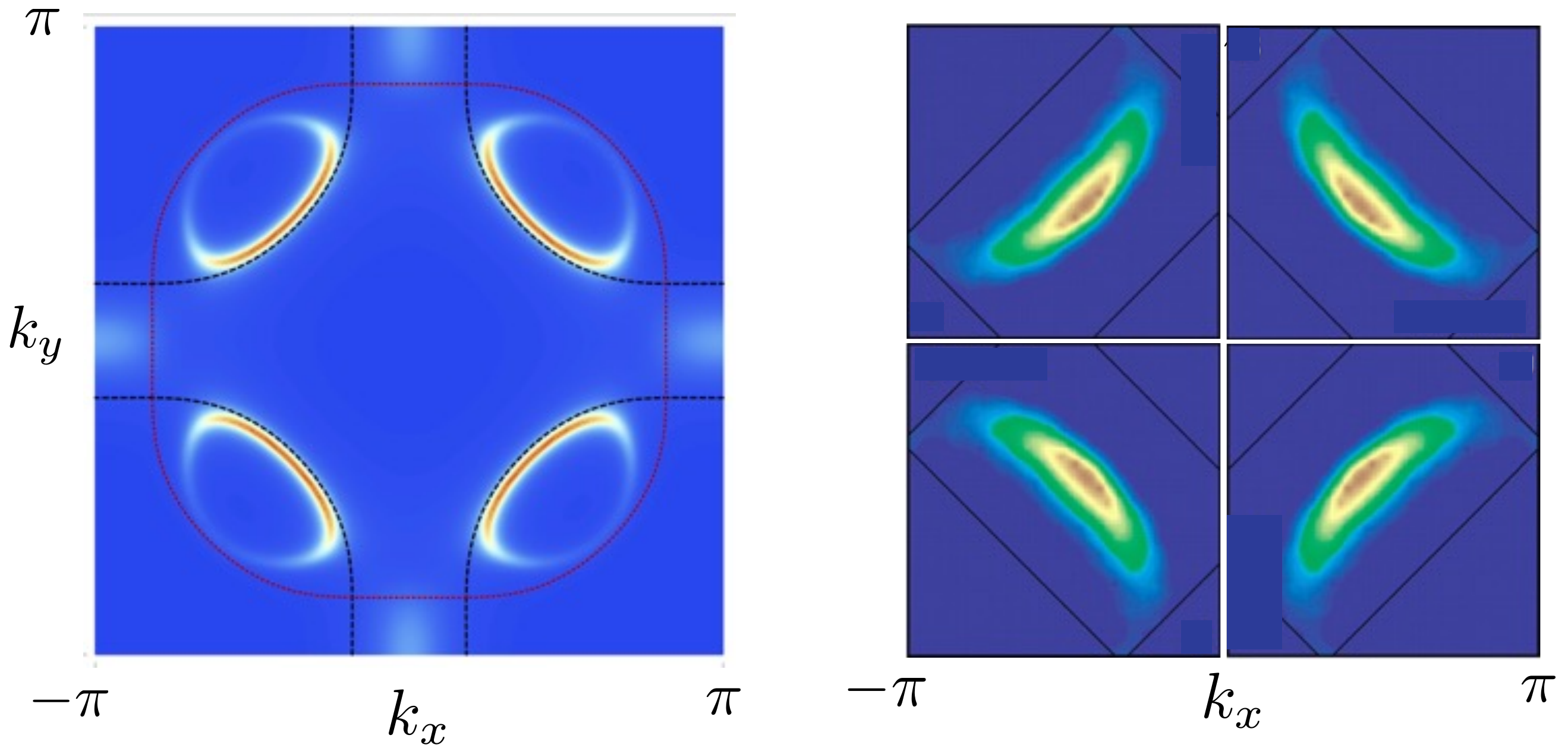}
\end{center}
\caption{On the left is the spectral function of the FL* state computed in Ref.~\citenum{YQSS10}. 
On the right are photoemission measurements of the pseudogap phase by Shen {\it et al.} in Ref.~\citenum{Shen05}.}
\label{flsdw}
\end{figure}

The spectral function corresponding to the above Green's function, along with comparison to experimental photoemission spectra, is shown in the left panel of Fig.~\ref{flsdw}. Note that the theoretical `small' pocket Fermi surfaces have an area given by $p$, and not $(1+p)$. This gives rise to a violation of Luttinger's theorem, which is however allowed because of the presence of `topological-order' due to the background U(1) spin-liquid. However these are pockets of hole-like quasiparticles, which are gauge-neutral under the emergent U(1) field, $A_\mu$. 

The FL* is a conventional Fermi-liquid phase, as far as its transport, thermodynamic or spectral properties are concerned. However, as noted above, the subtle difference arises from the presence of topological order. The spectral  weight on the `back-side' of these pockets is suppressed strongly. It has been argued that the Fermi-arcs in the pseudogap regime could in fact be modeled as pockets whose back-sides are (almost) invisible \cite{YRZ06,PJ11}. However, clear signatures of the `back-sides' are currently lacking in experiments.

\section{Connecting the high and low $T$ regimes}
\label{largeFS}

We will now connect our description of the FL* model in Section~\ref{flstar} to the charge order and superconducting instabilities 
presented in Section~\ref{pheno}.

For theoretical  purposes, and also with the aim of reviewing part work, it is useful to first present the instabilities of a conventional 
Fermi liquid state in Section~\ref{fl}. It turns out that these instabilities lead to some discrepancies with experimental observations. 
We will then argue in Section~\ref{flstarinst} that these discrepancies are naturally resolved by replacing the Fermi liquid by the FL* phase as the parent
phase of the orderings at low $T$.

\subsection{Ordering instabilities of a Fermi liquid with antiferromagnetic correlations}
\label{fl}

\begin{figure}[h]
\begin{center}
\includegraphics[width=2in]{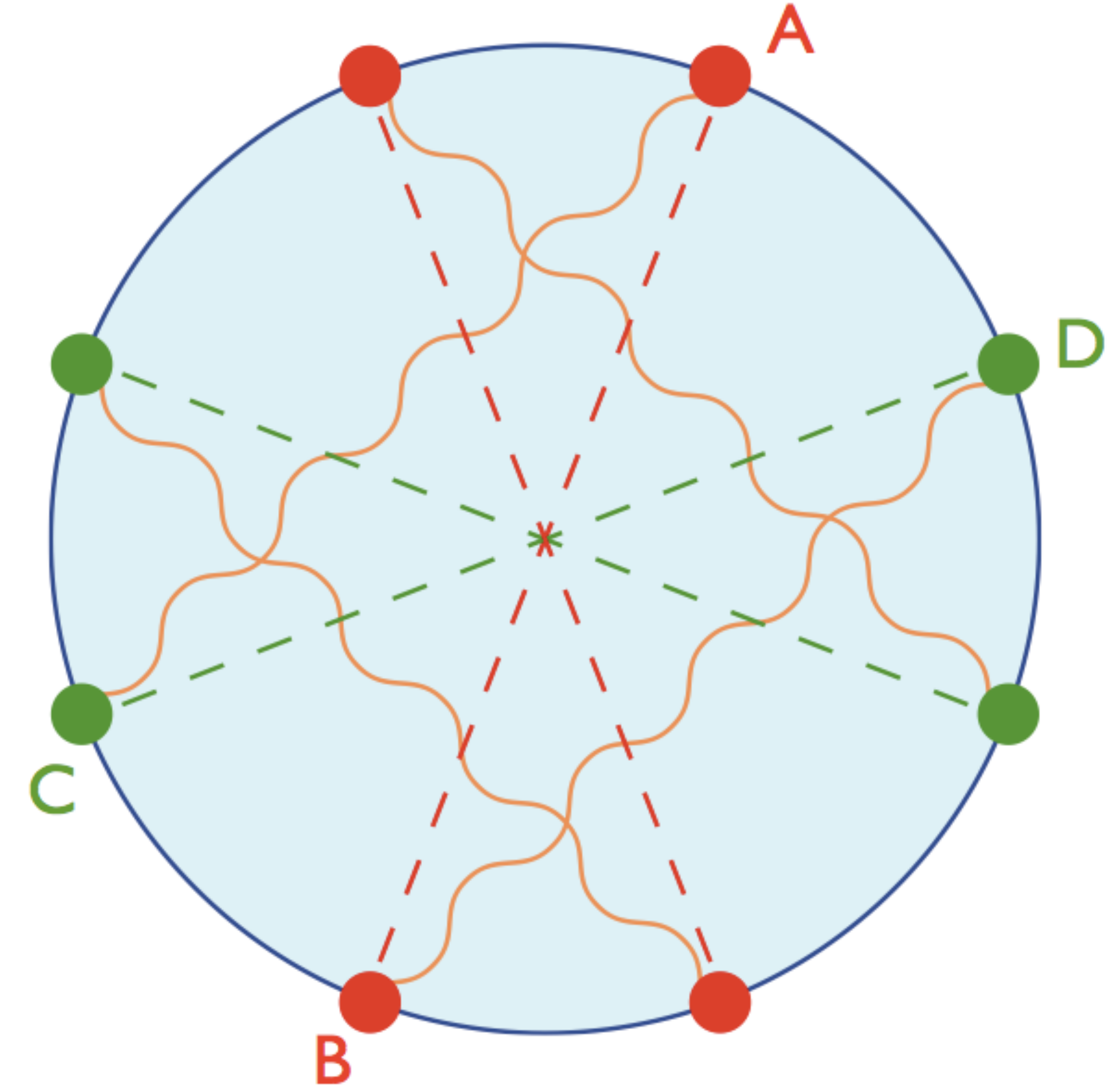}
\end{center}
\caption{A cartoon of a large Fermi surface, with the filled electronic states in blue. The red and green circles represent the `hot-spots' connected by $\K$ (shown as wavy lines). }
\label{hs}
\end{figure}

In this subsection, we return to the soft-spin theory in Eq.~(\ref{sdw}) for the properties of a Landau Fermi liquid in the presence
of amplitude fluctuations of a SDW order. Starting from Eq.~(\ref{sdw}) it is useful to focus on the low energy theory for fermions near 
the `hot spots' on the Fermi surface, shown in Fig.~\ref{hs}.
Such a Lagrangian is given by,
\beq
{\cal{L}}_{sdw} &=& {\cal{L}}_f + {\cal{L}}_\vp + {\cal{L}}_{f\vp},\nonumber\\
{\cal{L}}_f &=& \psi_{1p}^{\dagger m}(\partial_\tau - i\v_1^m\cdot\nabla)\psi_{1p}^m + \psi_{2p}^{\dagger m}(\partial_\tau - i\v_{2}^m\cdot\nabla)\psi_{2p}^m,\nonumber\\
{\cal{L}}_{f\vp} &=& \frac{\lambda}{\sqrt{N_f}} ~\vec\vp\cdot (\psi_{1p}^{\dagger m}\vec\sigma_{pp'}\psi_{2p'}^m + \psi_{2p}^{\dagger m}\vec\sigma_{pp'}\psi^m_{1p'}),\nonumber\\
\label{hstheory}
\eeq
and where ${\cal{L}}_\vp$ part of the action was already written down in Eq.\ref{SF}. In the above, $\v_1^m,~\v_2^m$ represent the velocities at the hot-spots labelled `A', `C', or, `B', 'D' within each hot-spot pair (denoted by `$m$').

Before proceeding any further, let us first digress for a bit and discuss an interesting (exact) symmetry associated with the exchange-interactions \cite{IAPWA88,LNW06,EF88}, that also manifests itself as an emergent symmetry \cite{MMSS10b} of the hot-spot theory. To start with, consider just the nearest neighbor Heisenberg term,
\beq
H_{J}=\sum_{i<j}J_{ij}\vec{S}_i\cdot\vec{S}_j,
\eeq
where $\vec{S}_i$ is expressed in terms of the underlying electronic operators in the usual way. If we, for instance, make the transformation, $c_{i\uparrow}^\dagger\rightarrow c_{i\downarrow}$, $c_{i\downarrow}^\dagger\rightarrow -c_{i\uparrow}$, then $S_i^z=n_{i\uparrow}-n_{i\downarrow}\rightarrow S_i^z$ and similarly for $S_i^\pm$. The above particle-hole transformation carried out on every site leaves the Hamiltonian invariant.  

It is possible to generalize the above transformations to arbitrary SU(2) rotations in particle-hole space. In order to do this, let us introduce the Nambu spinors,
\beq
\Psi_{i\uparrow}=\left(\begin{array}{c} c_{i\uparrow} \\ c_{i\downarrow}^\dagger \end{array} \right), ~~\Psi_{i\downarrow}=\left(\begin{array}{c} c_{i\downarrow} \\ -c_{i\uparrow}^\dagger \end{array} \right).
\eeq 
Then, $H_J$ can be re-expressed as,
\beq
H_J=\frac{1}{8}\sum_{i<j} J_{ij} \bigg( \Psi_{i\alpha a}^\dagger \vec\sigma_{\alpha\beta} \Psi_{i\beta a}\bigg)\cdot \bigg(\Psi_{j\gamma b}^\dagger \vec{\sigma}_{\gamma\delta}\Psi_{j\delta b} \bigg),
\eeq
where $a, b$ are Nambu indices. It is now a trivial matter to see that $H_J$ is invariant under independent SU(2) pseudospin transformations acting on each lattice site of the form $\Psi_{i\alpha a}\rightarrow U_{i,ab}\Psi_{i\alpha b}$.

We note that this pseudospin symmetry is broken explicitly by terms involving Coulomb repulsion and a finite chemical potential (except when we are at a fine-tuned, half-filled state). However, an interesting feature of the low-energy critical theory in Eq.\ref{hstheory} is that the SU(2) symmetry re-emerges. To see this, note that in Eq.\ref{hstheory}, if we carry out the following transformations,
\beq
\psi_{1\uparrow}^\dagger\rightarrow \psi_{1\downarrow},~\psi_{1\downarrow}^\dagger\rightarrow -\psi_{1\uparrow},~\psi_{2\uparrow}^\dagger\rightarrow \psi_{2\downarrow},~\psi_{2\downarrow}^\dagger\rightarrow -\psi_{2\uparrow}, \vec\vp\rightarrow\vec\vp,
\eeq
then ${\cal{L}}_{sdw}$ is invariant because,
\beq
\psi_{1\uparrow}^\dagger (\vec{v}_1\cdot\nabla) \psi_{1\uparrow}\rightarrow \psi_{1\downarrow}^\dagger (\vec{v}_1\cdot\nabla) \psi_{1\downarrow} = \psi_{1\downarrow} (\vec{v}_1\cdot\nabla) \psi_{1\downarrow}^\dagger = \psi_{1\downarrow}^\dagger (\vec{v}_1\cdot\nabla) \psi_{1\downarrow},
\eeq
where we have used integration-by-parts. The remaining terms can be simplified similarly. Note that it is crucial for the  curvature-terms to be absent in order for this symmetry to manifest itself. The SDW theory with 4 pairs of hot-spots therefore has an emergent [SU(2)]$^4$ symmetry, in addition to the usual spin SU(2) rotation symmetry.

\begin{figure}[h]
\begin{center}
\includegraphics[width=4.5in]{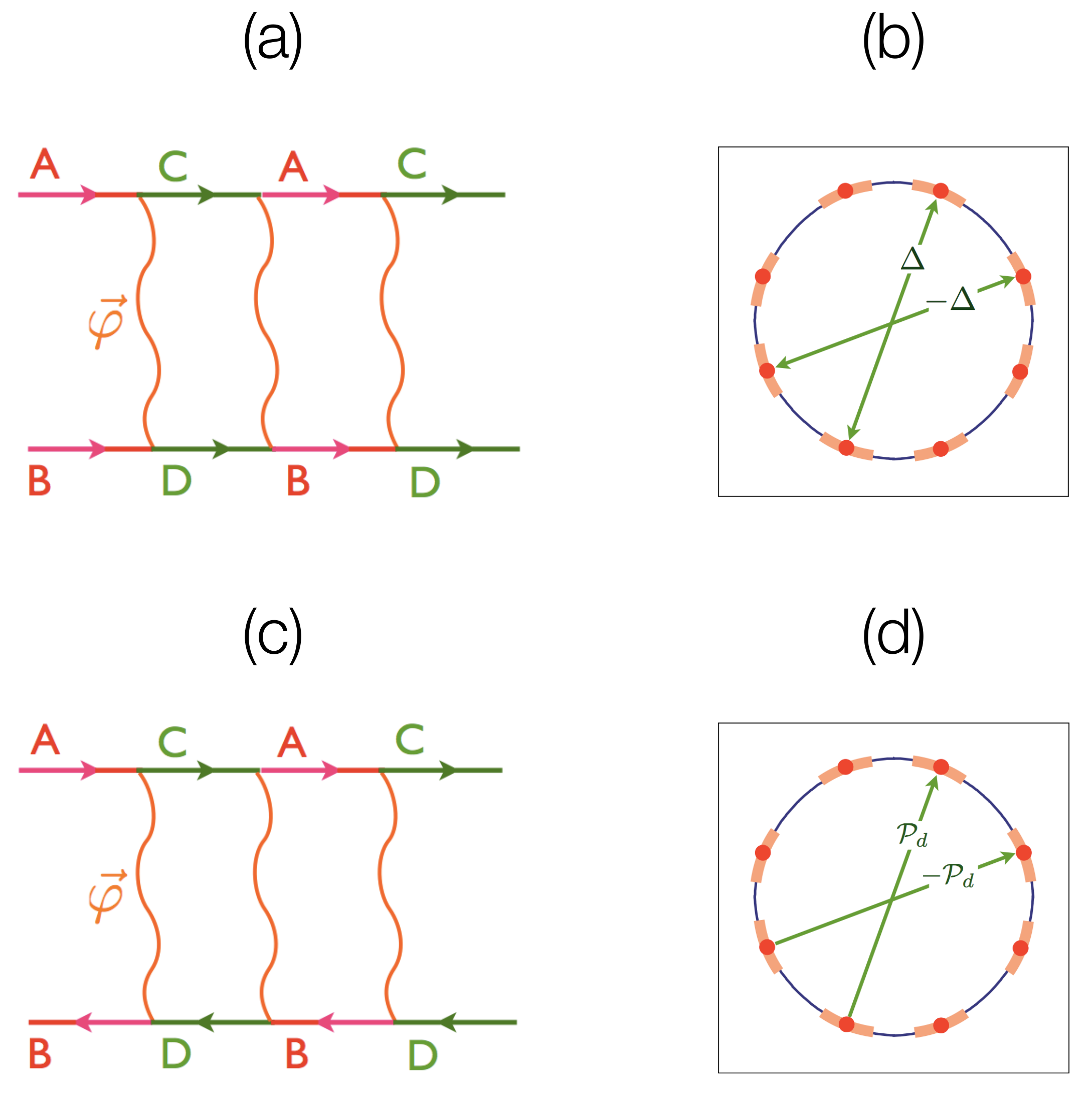}
\end{center}
\caption{(a) `Cooperon'-diagram in the particle-particle channel leads to (b) $d$-wave pairing with a relative phase-shift between pairs `A-B' and `C-D'. (c) `Diffuson'-diagram in the particle-hole channel, obtained from (a) by carrying out the SU(2) transformation on only half of the hot-spot pairs leads to (d) $d$-form factor density wave with $\Q=2k_F$ and a relative phase-shift between pairs `A-B' and `C-D'.}
\label{su2sym}
\end{figure}

Returning to the problem of a Fermi-liquid in the presence of SDW fluctuations, we recall a result that has now been known for almost three decades. The spin-fluctuations give rise to a pairing-glue in the $d_{x^2-y^2}$ channel \cite{DS86,CMV86}, as can be seen by explicitly evaluating the `Cooperon' diagram (Fig.\ref{su2sym}(a), (b)). The underlying reason for the attraction in the sign-changing $d$-wave channel can be traced back to the BCS gap-equation for the pairing gap $\D_\k$:
\beq
\D_\k=-\sum_\p \frac{\chi(\p-\k)}{2E_\p}~\D_\p.
\eeq
$\chi(\p-\k)$ is the SDW susceptibility, peaked at $\p-\k=\K$, the momentum carried by the SDW order-parameter and $E_\p=\sqrt{\e_p^2+\D_\p^2}$ is the dispersion for the Bogoliubov quasiparticles. In order for the above equation to have a non-trivial solution, $\tn{sgn}(\D_k)=-\tn{sgn}(\D_\p)$, whenever $\p-\k=\K$.

We can now use the SU(2) symmetry to carry out a particle-hole transformation on only the electrons in the vicinity of `B' and `D' (Fig.\ref{hs}). This then suggests that the same antiferromagnetic fluctuations would give rise to `pairing' in the particle-hole channel as well, as shown by the `diffuson' diagram in Fig.\ref{su2sym}(c). The particle-hole condensate, $\langle c_{\k-\Q/2,\alpha}^\dagger c_{\k+\Q/2,\alpha}\rangle$, now carries a finite center-of-mass momentum, $|\Q|=2k_F$ and there is an internal $d_{x^2-y^2}$ form-factor,  ${\cal{P}}_d$ (Fig.\ref{su2sym} (d)). Note that this wavevector will always be oriented along a diagonal direction, i.e. it is of the form $\Q=(Q_0,\pm Q_0)$ where . 

For the hot-spot theory, the superconducting and density-wave instabilities are exactly degenerate. However, a finite curvature which breaks the SU(2) symmetry, will lift this degeracy. While the superconducting instability remains unaffected (since $\pm \k$ are always `nested'), the density-wave instability becomes sub-leading to superconductivity.

The most natural question, in light of our discussion so far, is as follows: {\it Is there any relation between the density-wave instability found above in the vicinity of a SDW quantum-critical point, to the state unveiled in the underdoped cuprates?} 

\subsection{Relation to BDW in cuprates}

\begin{figure}[h]
\begin{center}
\includegraphics[width=2.1in]{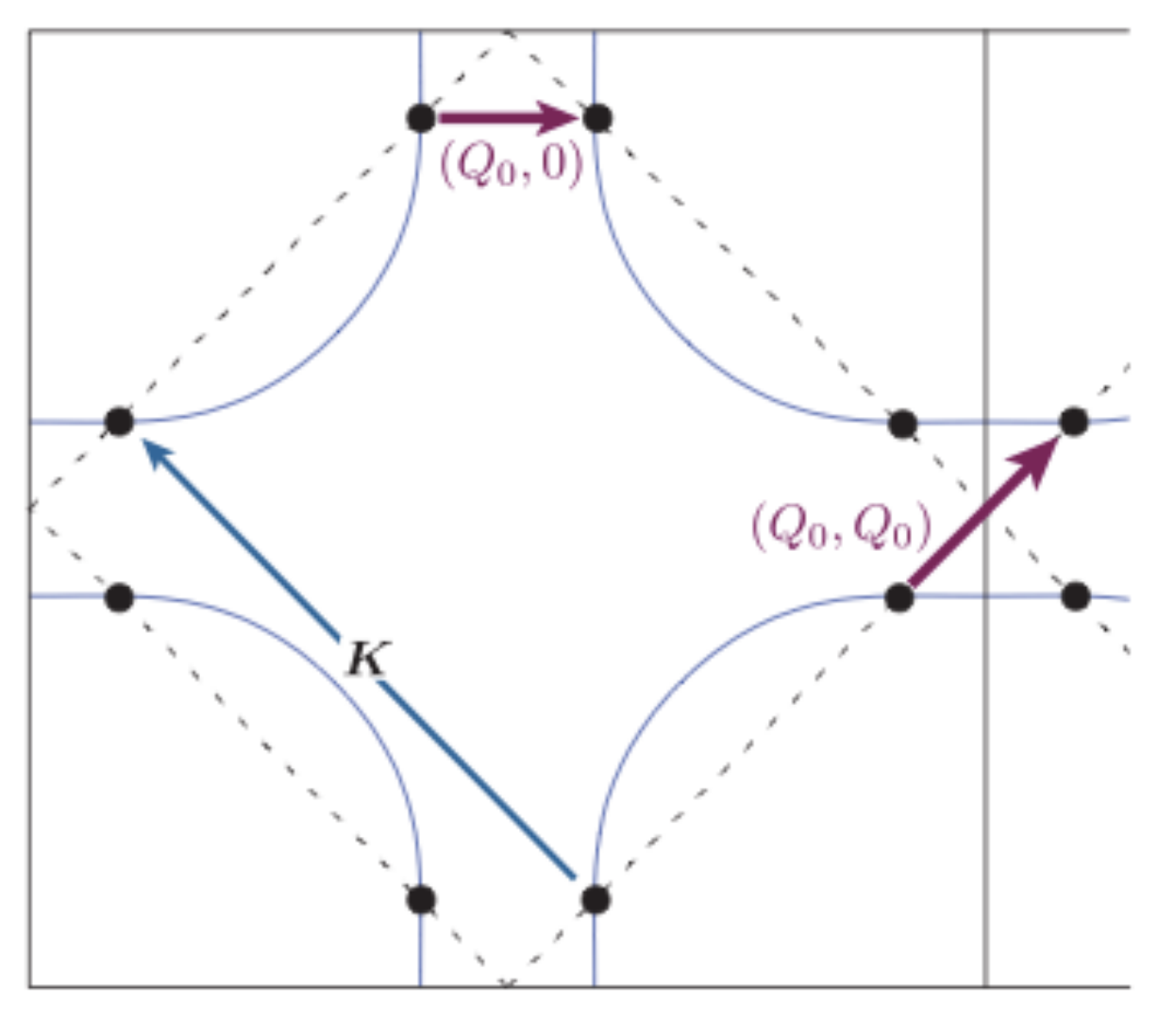}~\includegraphics[width=2.3in]{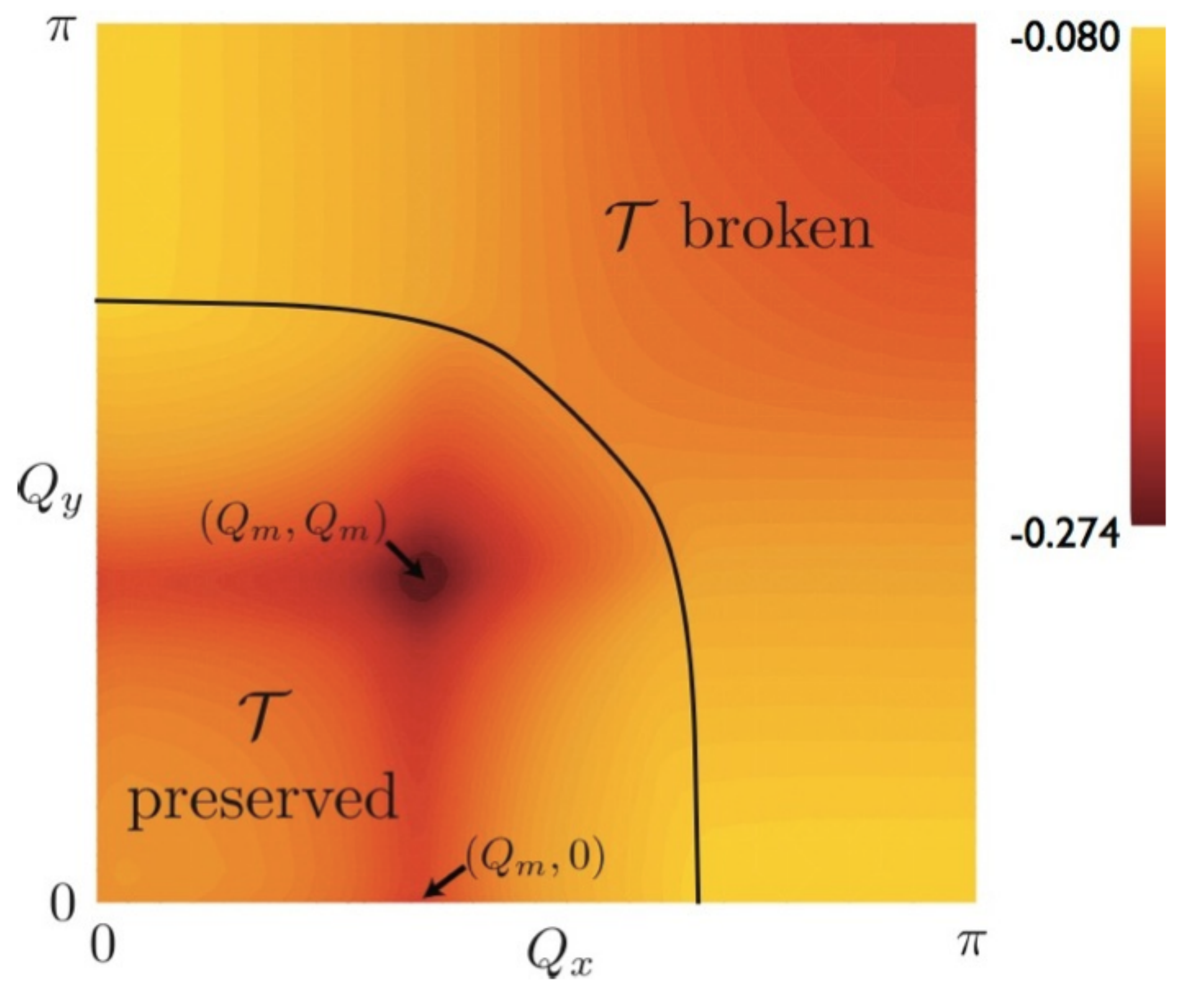}
\end{center}
\caption{Left: A cuprate-like Fermi surface with $t_1=1.0,~t_2=-0.32,~t_3=0.128,~\mu=-1.11856$. Black circles represent the hot-spots. The arrows represent wavevectors $\Q_a^x=(Q_0,0)$ and $\Q_b=(Q_0,Q_0)$ corresponding to BDW-a and BDW-b. Right: $\lambda_\Q$ as a function of $\Q$, showing the global minimum at $\Q_b$ and two `valleys' of local minima extending from $\Q_b$ to $\Q_a^x,~\Q_a^y$. Another local minimum appears close to $\Q=(\pi,\pi)$, which also breaks time-reversal symmetry and represents the `staggered-flux' state \cite{LNW06}. Adapted from 
Ref.~\citenum{SSRLP13}}
\label{tJ}
\end{figure}

Our aim is to investigate the nature of density-wave instabilities for the cuprate-like Fermi surface. A number of recent works \cite{SSRLP13,YWAC14,JSSS14,DCSS14b,AASS14a} have analyzed these instabilities for the full Fermi surface (Fig.\ref{tJ}a), starting with the work in Ref.~\citenum{SSRLP13}, which analyzed the form-factors and wavevectors of the (sub-)leading instabilities within a Hartree-Fock computation for $H=H_t+H_J$. The lowest eigenvalue, $\lambda_\Q$, of the particle-hole T-matrix determines the structure of the charge-order and is shown in Fig.\ref{tJ} as a function of $\Q$. 

Remarkably, the leading density-wave instability occurs for a wavevector $\Q_b=(Q_0,\pm Q_0)$,  as shown in Fig.\ref{tJ}a. The form-factor for this BDW is purely $d$-wave, as required by symmetry. Moreover, $\Q_b$ tracks the diagonal hot-spot wavevector over a range of doping. Thus even for the lattice Hamiltonian, and far away from any putative AFM QCP, the leading instability in the particle-hole channel corresponds to the instability that we had obtained earlier in the hot-spot theory. The one important difference however is that this instability is not degenerate with the $d$-wave superconducting instability, as expected in the absence of the exact SU(2) symmetry. The eigenvalues of the T-matrix in the particle-particle channel were also analyzed \cite{DCSS14}, allowing for the possibility of a finite center-of-mass momentum, $\Q$ (i.e. a pair-density wave, PDW, state \cite{EBPDW,YWDAAC14}). However, the leading weak-coupling instability was always to a $d$-wave SC with $\Q=0$. It has been argued recently that there could be a strong-coupling instability to a PDW state, without reference to any microscopic model, arising from an effect analogous to `Ampere's law' \cite{PAL14}. 

Coming back to our discussion of $\lambda_\Q$, one can also notice the `valleys' of local minima that extend from $\Q_b$ to $\Q_a^x=(Q_0,0)$ and $\Q_a^y=(0,Q_0)$. The form-factor for these BDWs predominantly have a $d$-wave component with a small $s, s'-$component. Notice that the $\Q_a^{x,y}$ are oriented along the Cu-O directions and are therefore quite similar, at least qualitatively, to the states observed in the experiments. However, while here the wavevectors connect the hot-spots, it is unclear to what extent this is the case experimentally across different families of the cuprates.

\begin{figure}[h]
\begin{center}
\includegraphics[width=5in]{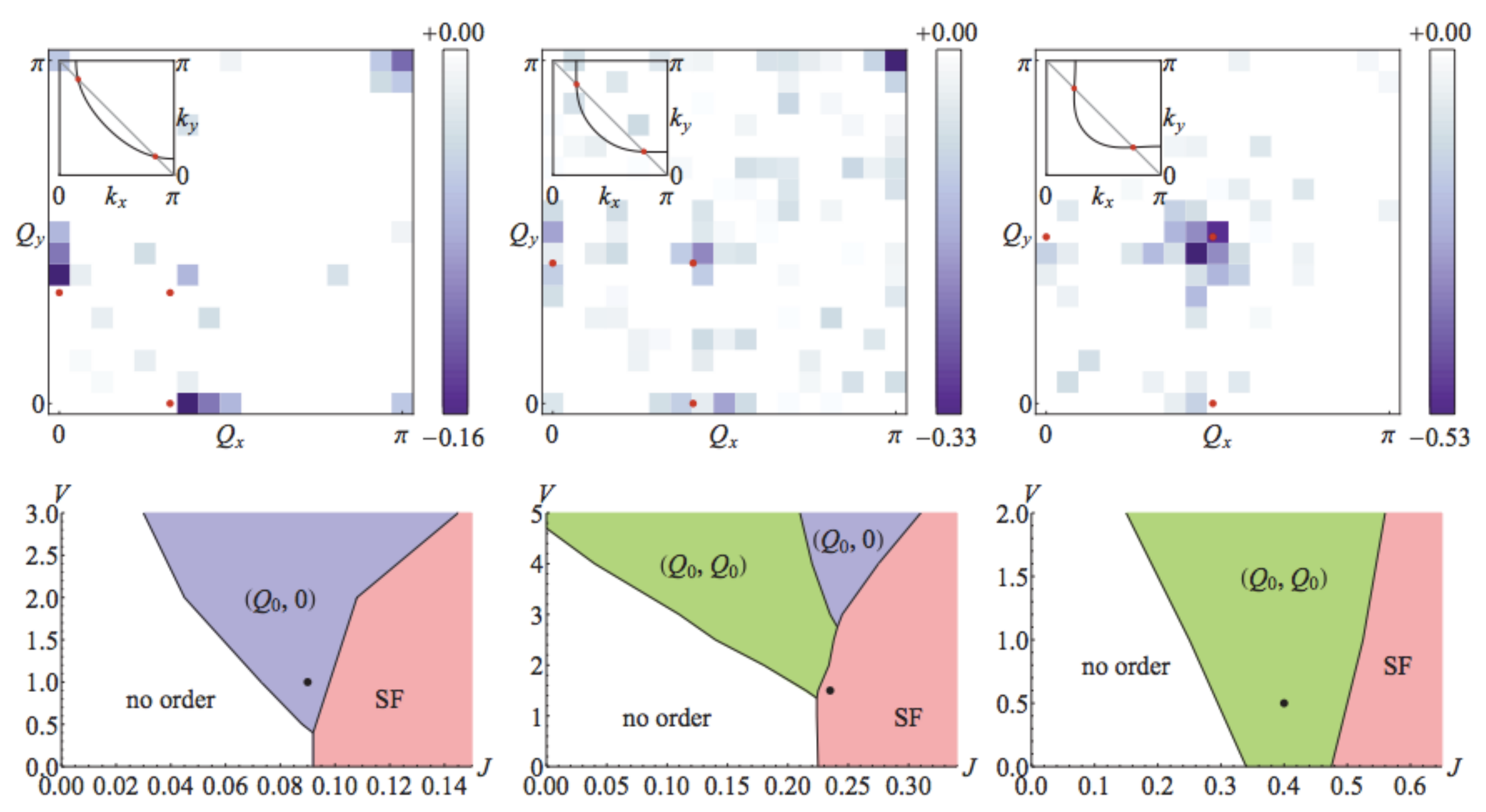}
\end{center}
\caption{Top row: Gain in variational energy for ordering at wave vector $\Q$; few representative parameter sets have been shown, corresponding to axial BDW, staggered flux and diagonal BDW respectively. The red dots denote the hot-spot wavevectors. Bottom row: Approximate diagrams of the leading instability in the particle-hole channel, as a function of $J$ and $V$. Adapted from Ref.~\citenum{AASS14a}. }
\label{vmc}
\end{figure}

The above computation was somewhat rudimentary---it only analyzed the quadratic instabilities of the $t$-$J$ model (without projection). A subsequent study \cite{AASS14a} carried out a variational Monte-Carlo computation for the leading instabilities of the $t$-$J$ model, with an infinite on-site Hubbard $U$ (implemented using Gutzwiller projection). Their results are shown in Fig.\ref{vmc}. Interestingly, there were regimes in parameter space where for at least some Fermi surface geometries, the BDW state with $\Q_a$ was found to be the leading instability. However, this required somewhat large nearest neighbor Coulomb repulsion ($V\gg J$); furthermore, given the large coupling and the numerical projection, we cannot be certain that this instability is arising from a Fermi liquid with a large Fermi surface.

The above analysis leads to two key insights, which is responsible for the diagonal BDW with a $d$-form factor being the leading density-wave instability: ({\it i\/}) the remnant of the SU(2) pseudospin symmetry, arising from the exchange interactions, after it has been broken explicitly is still strong enough, and, ({\it ii\/}) a large joint density of states is available for hot-spots along the diagonal wave vector, which favors the diagonal BDW. One possible way around this is to preferentially suppress the density of states in the antinodal regions. 

A natural ingredient that suppresses density of states at the antinodes is superconductivity. However, since the BDW correlations onset above $T_c$, one can only allude to superconducting fluctuations. Indeed, pairing fluctuations in a metal suppress the diagonal BDW much more strongly compared to the axial BDW at low enough temperatures \cite{DCSS14,CPKE14}. On the other hand, experiments have detected the presence of these pairing fluctuations \cite{MRCK05} to a relatively narrow window above $T_c$, while the onset of BDW occurs significantly above $T_c$. Moreover, this approach still does not rule out why the diagonal BDW has never been detected in the cuprates at high enough temperatures.

Therefore, one needs a different mechanism to suppress the antinodal density of states. One other possibility is to use long-range commensurate AFM to gap out the antinodes completely \cite{AK14,ATSS14}. The instabilities of a reconstructed Fermi surface with hole-pockets (Fig. \ref{sdw}) does give rise to a density-wave with the correct wavevectors, but with a form-factor that is predominantly $s'$ \cite{ATSS14}. However, this can not be an explanation for the charge-ordering phenomenon in the non-La-based cuprates where the AFM correlation length is less than one lattice constant. 

It is then perhaps appropriate to investigate the instabilities of a `normal' state that already has a well-developed pseudogap, without any symmetry-breaking, i.e. one should investigate the instabilities of a state that has Fermi-`arcs', and not a large Fermi surface. This naturally leads us to investigate the instabilities of the FL* introduced in Section~\ref{flstar}.

\subsection{Ordering instabilities of FL*}
\label{flstarinst}

Computing the ordering instabilities of the FL* metal is a far more complicated task than that for the Fermi liquid described
in Section~\ref{fl}. Apart from instabilities associated with particle-particle and particle-hole excitations near the Fermi surface,
we also have to consider the instabilities associated with the gauge sector. For the U(1)-FL* case, the latter contain monopole
tunneling events which necessarily lead to confinement and broken symmetries at low temperatures. There has been partial progress
on these difficult issues in the literature, but we shall not review them here. Instead we shall focus only on the 
charge ordering instabilities associated with the `small' Fermi surface, in close analogy to the analyses of Section~\ref{fl}.

\begin{figure}[h]
\begin{center}
\includegraphics[width=4.7in]{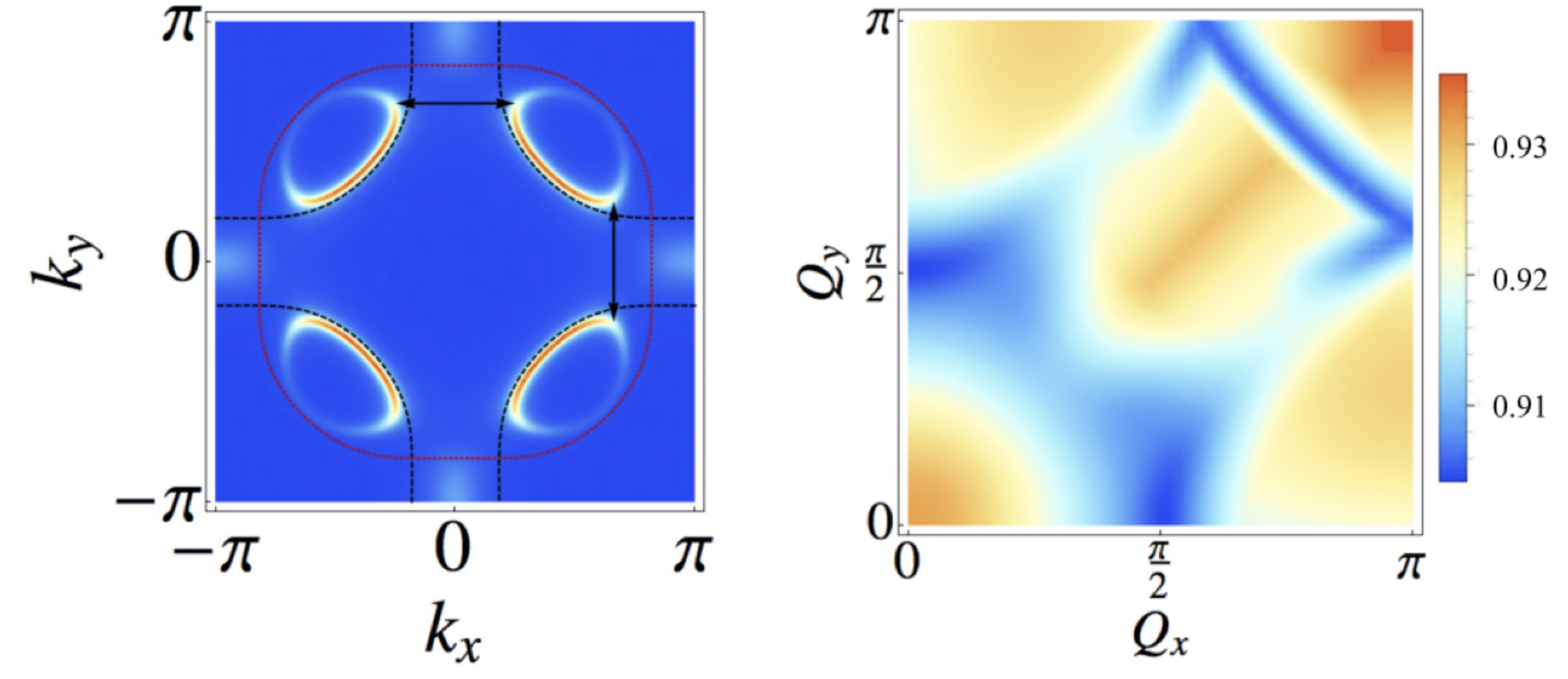}
\end{center}
\caption{Left: The spectral function for a FL* state with the same FL parameters as in Fig.~\ref{tJ}. The other parameters are $\tilde{t}_0=-0.5t_1,~ \tilde{t}_1=0.6t_1,~\lambda=0.75t_1$ (Eq.\ref{disp}). The black arrows denote the wave vectors ($\Q_a$) of the leading density-wave instability. Right: $\lambda_\Q$ as a function of $\Q$, showing the global minimum at $\Q_a$. Adapted from Ref.~\citenum{DCSS14b}. }
\label{flsdwstar}
\end{figure}

The surprising outcome is that even a somewhat rudimentary RPA level analysis of the density-wave instabilities of the FL* immediately leads to a state with $d$-form factor and wavevectors along the axial directions, as displayed in Fig.~\ref{flsdwstar}. Moreover, the wavevector is related to a geometric property of the underlying Fermi surface --- it is determined by the distance between the tips of the pockets. Hence, as the hole-doping increases, the magnitude of the wavevector decreases, consistent with the experimental results on the non-La-based cuprates \cite{BB13}.

The above discussion seems to suggest that the $d$-form factor BDW with the axial wavevectors, observed experimentally at low temperatures, arises most naturally out of a parent high-temperature FL* phase. However, the plain vanilla FL* state has hole pockets with nonzero (but small) spectral weight on the backside. While the photoemission results of Ref.\citenum{PJ11} have been argued to be approximately consistent with this, clear signatures of the backside continue to be elusive. A promising future direction for experiments should be to employ probes which do not involve adding or removing an electron from the sample (which are sensitive to the quasiparticle residue), but instead measure the $2k_F$ response corresponding to the `small' Fermi surfaces directly using Friedel oscillations, the Kohn anomaly, or ultrasonic attenuation.

\begin{figure}[h]
\begin{center}
\includegraphics[width=4.7in]{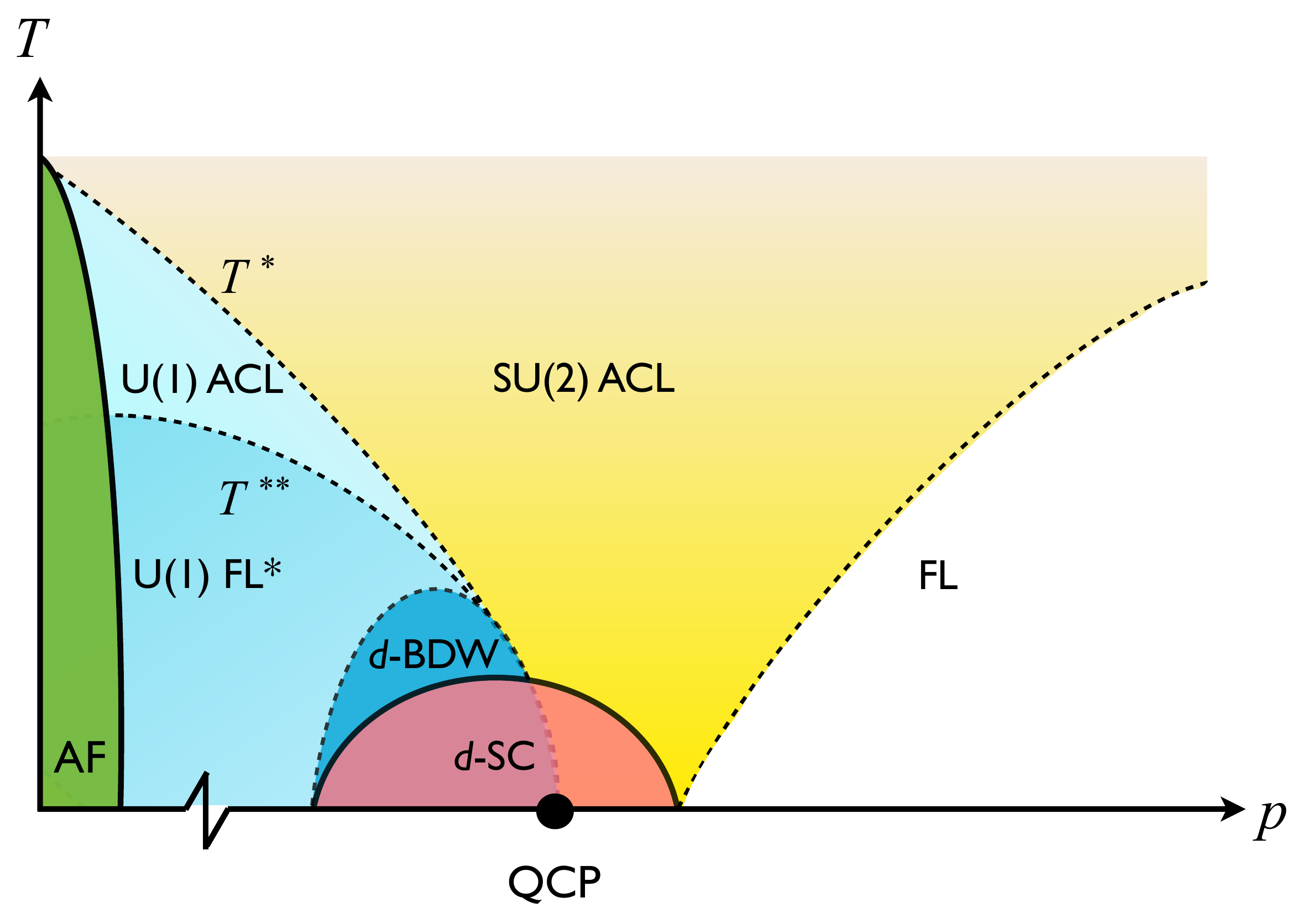}
\end{center}
\caption{Our proposed global phase diagram for the cuprates, adapted from Ref.~\citenum{DCSS14c}.
The present review has focused on the regime to the left of the Higgs quantum critical point (QCP), with the U(1) FL* and its instability
to the $d$-BDW. The Higgs QCP describes a phase transition from the U(1) ACL to the SU(2) ACL which is accompanied by a 
change from `small' to `large' Fermi surfaces.}
\label{su2}
\end{figure}

In Section~\ref{flstar}, both prescriptions of arriving at the FL* relied on starting from the antiferromagnetic insulator and doping it with charge-carriers. A natural question that now needs to be addressed is how to connect the metallic FL* phase with a `small' Fermi surface to the `large' Fermi surface metal. As discussed recently \cite{SS09,DCSS14c}, it is possible to describe this in terms of a Higgs' transition, that does not involve any broken symmetries and yet reconstructs the Fermi surface. However the transition proceeds via an intermediate non Fermi-liquid phase which was dubbed an SU(2) {\it algebraic charge-liquid} (ACL). Our view of the proposed phase diagram is reproduced in Fig.~\ref{su2} for the sake of completeness. We refer the interested reader to Ref.~\citenum{DCSS14c} for a more complete discussion of the Higgs' transition and a comparison to some of the associated phenomenology in the cuprates. 

\section{Outlook}
\label{outlook}

The main purpose of this review was to address a number of pertinent questions regarding the nature of the newly discovered charge-order in the (non-La-based) underdoped cuprates and to explore its connection with the high temperature pseudogap phase. Let us revisit the questions that we had raised in the introduction one by one, in order to summarize our current understanding and provide a brief outlook:
\begin{itemize}
\item The $d$-form factor BDW likely arises as a low-temperature instability of a fractionalized Fermi-liquid (FL*) with `small' Fermi surfaces of hole-like quasiparticles \cite{DCSS14b}. While the FL* resembles a usual Fermi-liquid in terms of its transport and thermodynamic properties, it violates Luttinger's theorem on the area enclosed by the Fermi surface, and this is closely linked to the presence of emergent 
excitations belonging to a gauge sector \cite{TSMVSS04}. 
\item The FL* has pocket Fermi surfaces with low intensity on the `back sides' because of small (but non-zero) 
quasiparticle residue \cite{RK08,YQSS10}. This is presumably the reason only `Fermi arcs' have been detected by photoemission so far. 
Fermi surface probes which don't involve adding or removing an electron could provide detection of the full pockets.
\item The modulation wave vector of the BDW is along the axial directions and is determined by the distance between the `tips' of the pockets \cite{comin13,neto13}. The $d$-form factor arises as a result of the remnant SU(2) particle-hole symmetry associated with the exchange interactions \cite{MMSS10b}, that maps $d$-wave superconductivity to $d$-form factor BDW.
\item The non-mean field nature of the onset of BDW and SC correlations is governed by the angular fluctuations of a six-component order-parameter. A phenomenological non-linear $\sigma$-model for the multicomponent order-parameter provides a quantitative description for the onset of charge-order intensity as well as the diamagnetic response in the pseudogap regime of the underdoped cuprates \cite{LHSS14,LHSS14b}. 
\item 
A large Fermi surface in the presence of long-range (and nearly bidirectional) charge order leads to a nodal electron pocket \cite{HS11} flanked by recently predicted small hole pockets \cite{AADCSS14}. These are likely responsible for the observed quantum oscillations \cite{LThole14}. 
At low temperatures and high fields, we expect the U(1) FL* to undergo a crossover into a confined phase with charge order; it is possible that
the nodal Fermi surface region is insensitive to confinement and so has a pocket structure similar to that obtained from the large Fermi surface
computation in Ref.~\citenum{AADCSS14}. It is clear that this crossover to confinement is an important topic for future research.
\item A candidate theory \cite{DCSS14c} for the optimal-doping QCP is associated with a Higgs' transition, without any broken symmetries, from a SU(2) ACL to a U(1) ACL.
\end{itemize}

\section{Acknowledgements}

We thank our collaborators Andrew Achkar, 
Andrea Allais, Johannes Bauer, S\'eamus Davis, Kazuhiro Fujita, Mohammad Hamidian, David Hawthorn, Lauren Hayward, Rolando La Placa, Roger Melko, 
Max Metlitski,  Matthias Punk, and Yang Qi.
We also thank Andrey Chubukov,  Martin Greven, T. Senthil and Louis Taillefer for valuable discussions. 
D.C. is supported by the Harvard-GSAS Merit fellowship.
This research was supported by the NSF under Grant DMR-1360789, the Templeton foundation, and MURI grant W911NF-14-1-0003 from ARO.
Research at Perimeter Institute is supported by the Government of Canada through Industry Canada 
and by the Province of Ontario through the Ministry of Research and Innovation. 

\bibliography{cuprates}
\end{document}